\documentclass[11pt]{article}
\usepackage{cite,a4wide}
\newcommand{\zb}{\mbox{$\rm Z^0$}}

\newcommand{\polem}{\mbox{${\cal P}^-$}}
\newcommand{\pmi}{\polem}
\newcommand{\polep}{\mbox{${\cal P}^+$}}
\newcommand{\ppl}{\polep}
\newcommand{\pol}{\mbox{${\cal P}$}}
\newcommand{\bqbar}{\mbox{$\rm\bar{b}$}}
\newcommand{\bbbar}{\mbox{$\rm b\overline{b}$}}
\newcommand{\ccbar}{\mbox{$\rm c\overline{c}$}}
\newcommand{\ffbar}{\mbox{$\rm f\overline{f}$}}
\newcommand{\bo}{\mbox{$\rm B^0$}}
\newcommand{\bobar}{\mbox{$\rm \bar{B}^0$}}
\newcommand{\bplus}{\mbox{$\rm B^+$}}
\newcommand{\bs}{\mbox{$\rm B_s$}}
\newcommand{\ds}{\mbox{$\rm D_s$}}
\newcommand{\dm}{\mbox{$\Delta m$}}
\newcommand{\dms}{\mbox{$\Delta m_{\rm s}$}}
\newcommand{\acs}{\mbox{$a_{\rm cos}$}}
\newcommand{\asn}{\mbox{$a_{\rm sin}$}}
\newcommand{\jpsi}{\mbox{$\rm J/\psi$}}
\newcommand{\ks}{\mbox{$\rm K^0_s$}}
\newcommand{\pppm}{\mbox{$\rm\pi^+\pi^-$}}
\newcommand{\pzpz}{\mbox{$\rm\pi^0\pi^0$}}
\newcommand{\pppz}{\mbox{$\rm\pi^+\pi^0$}}
\newcommand{\dedx}{\mbox{${\rm d}E/{\rm d}x$}}
\newcommand{\bratio}[2]{\mbox{${\cal B}(#1\rightarrow #2)$}}
\newcommand{\fsig}{\mbox{$f_{\rm sig}$}}
\def\ifmath#1{\relax\ifmmode #1\else $#1$\fi}%
\newcommand{\ALR}{\mbox{$A_{\rm {LR}}$}}
\newcommand{\cAe} {\mbox{$\cal A_{\rm e}$}}

\newcommand{\cAq} {\mbox{$\cal A_{\rm q}$}}

\newcommand{\cAb} {\mbox{$\cal A_{\rm b}$}}

\newcommand{\swsqeffl}    {\sin^2\!\theta_{\rm{eff}}^\ell}

\newcommand{\Rb}   {\ifmath{R_{\mathrm{b}}}}

\newcommand{\Rbz}  {\ifmath{R_{\mathrm{b}}^0}}

\newcommand{\MZ}      {m_{\mathrm{Z}}}

\newcommand{\MT}      {m_{\mathrm{t}}}
\newcommand{\GZ}      {\Gamma_{\mathrm{Z}}}
\newcommand{\Afb}     {A_{\mathrm{FB}}}
\newcommand{\Afbb}    {A_{\mathrm{FB}}^b}

\newcommand{\Ghad}       {\Gamma_{\mathrm{had}}}

\newcommand{\Gll}        {\Gamma_{\ell}}

\newcommand{\GF}{G_{\mathrm{F}}}
\newcommand{\ee}{\mathrm{e}^+\mathrm{e}^-}

\newcommand{\cm}{\mathrm{cm}}

\def\pd{\phantom{-}}

\def\pd{\phantom{-}}

\newcommand{\PLB}[3] {Phys.~Lett.\ {\bf B#1} (#2) #3}
\newcommand{\PRL}[3] {Phys.~Rev.\ {\bf Lett.~#1} (#2) #3}

\newcommand{\NIM}[3] {Nucl.~Instrum.\ {Methods~{\bf #1}} (#2) #3}

\newcommand{\ZPC}[3] {Z.~Phys.\ {\bf C#1} (#2) #3}

\newcommand{\EPJ}[3] {Eur.~Phys.\ J.\ {\bf C#1} (#2) #3}
\newcommand{\etal} {et~al.}
\input epsf
\newcommand{\epostfig}[3]{
\begin{figure}[tbp]
\setlength{\epsfxsize}{1.1\hsize}
\hspace*{-0.05\hsize} \epsfbox{#1}
\caption{\label{#2}#3}
\end{figure}
}
\newcommand{\epostscale}[4]{
\begin{figure}[tbp]
\centering

\setlength{\epsfxsize}{#3\hsize}
\hspace*{-0.05\hsize} \epsfbox{#1}
\caption{\label{#2}#4}
\end{figure}
}
\begin{document}
{\flushright
DESY 99-157 \\
October 12, 1999 \\
}
\begin{center}
    \LARGE\bf\boldmath
Electroweak and CP violation physics\\
at a Linear Collider Z-factory\\
\end{center}
\vspace{5mm}
\bigskip

\begin{center}
\large R. Hawkings \\
DESY-FLC / Universit\"at Hamburg \\
K. M\"onig \\
DESY Zeuthen
\end{center}

\bigskip
\begin{abstract}
A future linear collider such as TESLA may be able to run on the \zb\
resonance with very high luminosity and polarised electron and positron
beams. The possibilities of measuring electroweak quantities with
high precision are investigated. Huge improvements with respect to
the present precision can be expected, especially for the asymmetries
$\ALR$ and $\cAb$ where beam polarisation can be exploited.
The very large sample of $\zb\rightarrow\bbbar$ events also
allows studies of various CP-violating b decays. The precision
achievable on the CKM unitarity triangle angles is comparable to 
experiments at b factories and future hadron colliders.
\end{abstract}

\section{Introduction}

The primary motivation for a next generation $\rm e^+e^-$ linear collider
is the study of collisions at centre of mass energies beyond 200\,GeV.
However, by adding an auxiliary positron source, it may be possible
to run with high luminosity (${\cal L}\sim 5\cdot 10^{33}\rm\,cm^{-2}s^{-1}$)
and polarised beams at the peak of the \zb\ resonance \cite{nick}.
This luminosity would allow the accumulation of $10^9$ hadronic
\zb\ decays in about 70 days running, opening up the possibility to repeat
the electroweak measurements currently performed at LEP and SLD with much
higher precision.

A sample of $10^9$ \zb\ decays represents a gain of about 100 over the
LEP1 data sample. However, with the beam polarisation and good b-tagging
capability at a linear collider, the more relevant comparison is with
SLD, where the gain will be more like a factor of 2000.
With beam polarisation,
the variable that is most sensitive to the effective weak mixing angle
$\swsqeffl$ is $\ALR$, the left-right asymmetry in \zb\ decays.
Using $\ALR$, SLD at SLAC was able to measure the weak
mixing angle to a better accuracy than any single observable at LEP with
less than a tenth of the luminosity \cite{lepew}. Although 
the present SLD analysis technique is limited by the precision of the
electron polarisation measurement, the availability of positron polarisation
at a linear collider allows the polarisation to be measured from cross
section asymmetries. Using a scheme first proposed by A.~Blondel for 
LEP \cite{alain}, external measurements are then only needed for polarisation 
differences.

This large data sample would also contain about $4\cdot 10^8$ b hadrons,
a large enough sample to study rare CP-violating b decays. The linear collider
offers a number of advantages over other facilities: the beam polarisation
allows a very good initial state tag from the b-quark forward-backward 
asymmetry, the data sample is larger than that from asymmetric $e^+e^-$ 
b factories,
whilst the environment is much cleaner than that at hadron colliders.
Competitive measurements of the angles of the CKM unitarity triangle may thus
be possible.

The precision attainable on such electroweak and CP-violation measurements
is studied in this paper. The basic assumptions used are described in 
Section~\ref{s:setup}, followed by discussions of the lineshape
parameters in Section~\ref{s:lines} and \Rb\ in Section~\ref{s:rb}.
The measurement of $\ALR$, including the Blondel scheme for the 
polarisation determination, is described in detail in Section~\ref{s:alr}, 
followed by $\cAb$ in Section~\ref{s:ab}. The potential for measurement
of the CKM unitarity triangle angles $\alpha$, $\beta$, $\gamma$ and 
$\delta\gamma$ is investigated in detail in Section~\ref{s:cp}. Finally,
the conclusions are given in Section~\ref{s:conc}.

\section{Data sample and assumptions}\label{s:setup}

The TESLA linear collider should be able to deliver a luminosity of
${\cal L} = 5 \cdot 10^{33} \cm^{-2} \rm{s}^{-1}$ at the \zb-pole with the
aid of an auxiliary positron source \cite{nick}. With this luminosity,
the sample of $10^9$ hadronic \zb\ (corresponding to
$30\,\rm fb^{-1}$) assumed in this paper
can be produced in about 70 days running---a small cost to the overall
physics program of the collider. The detector, trigger and data acquisition
must of course be able to cope with the trigger rate of 
200\,Hz from hadronic \zb\ decays.

It will be assumed that the electron and positron beams can be polarised with 
polarisations of $\pmi=\pm 80$\,\% and $\ppl=\pm 60$\,\%. This corresponds to
an equivalent polarisation of $\pol=\frac{\ppl+\pmi}{1+\ppl \pmi} \sim 95 \%$.
Additionally, the  polarisation should be switchable randomly from 
train to train.

The energies of both the electron and positron beams can be measured
with a spectrometer similar to the one currently being installed in LEP
\cite{lepspec}. Fast energy scans should be possible, 
so that the spectrometer can be calibrated relative to $\MZ$ from LEP.
With this scheme a precision of 1\,MeV might be reached. It should be
noted that the relevant parameter to correct the electroweak measurements is
the difference $\sqrt{s} - \MZ$, so that the error on $\MZ$ does not enter
when the spectrometer is calibrated relative to the \zb\ mass.

\section{Lineshape parameters}\label{s:lines}

The parameters that are measured at LEP in the scan around the \zb\
resonance are:
\begin{itemize}
\item the mass of the \zb\ $(\MZ)$
\item the width of the \zb\ $(\GZ)$
\item the hadronic peak cross section 
$(\sigma_0 \propto \frac{\Gll\Ghad}{\GZ^2})$
\item the ratio of the hadronic and leptonic partial widths
$(R_\ell = \frac{\Ghad}{\Gll})$
\end{itemize}
Since $\MZ$ is used to calibrate the spectrometer, its precision clearly cannot
be improved at TESLA. With the large \zb\ data sample,
the error on $\GZ$ is equal to the error on $\sqrt{s}$, so 1\,MeV is
within reach.

The error on the hadronic peak cross section is dominated by uncertainties
coming from the luminosity measurement, mainly from theory. It is possible
that with the high luminosity available at TESLA, processes with low
cross section but small uncertainties 
like $\ee \rightarrow \gamma \gamma$ can be used,
but here it will be assumed that no improvement on $\sigma_0$ will
be possible.

The partial width ratio $R_\ell$ 
receives substantial contributions from systematic uncertainties
due to the leptonic and hadronic event selection. The uncertainties from the
leptonic event selection, 
especially for muons, are mainly of a statistical nature,
so can be assumed to scale with the luminosity. The uncertainty
on the hadronic event selection will be assumed to stay at the best value
reached at LEP (0.05\%). A more careful study on this subject is in progress
\cite{marc}.

The interesting physics parameters that can be derived from the lineshape
parameters are
\begin{itemize}
\item the mass of the \zb\ $(\MZ)$
\item the strong coupling constant at the \zb-mass $(\alpha_s(\MZ^2))$
\item the radiative correction parameter normalising the strength of
the \zb-couplings to the fermions $(\Delta \rho)$
\item the number of light neutrino species $(N_\nu)$
\end{itemize}

The possible improvements in these parameters are summarised in Table
\ref{tab:line}. It can be seen that an interesting gain can be 
obtained for $\alpha_s(\MZ^2)$ and $\Delta \rho$. $N_\nu$ is completely
determined by $\sigma_0$, so no change is expected if the luminosity
determination cannot be improved.

\begin{table}
\begin{center}
\begin{tabular}[c]{c|c|c}
\hline
 & LEP \protect\cite{lepew} & TESLA \\
\hline
$\MZ$ & {$ 91.1867 \pm 0.0021\rm\,GeV$} & {$ \pm 0.0021\rm\,GeV$} \\
$\alpha_s(\MZ^2)$ & {$ 0.1212 \pm 0.0034 $} & {$ \pm 0.0016$} \\
$\Delta \rho$ & {$ (0.42 \pm 0.12 ) \cdot 10^{-2}$} 
& {$ \pm 0.05\cdot 10^{-2}$}  \\
$N_\nu$ & {$ 2.994 \pm 0.0011 $} & {$ \pm 0.0011 $} \\
\hline
\end{tabular}
\end{center}
\caption{Possible improvements in the physics quantities derived from the
\zb-lineshape after high luminosity \zb-running at TESLA.
For $\alpha_s$ and $\Delta \rho, N_\nu=3$ is assumed.}
\label{tab:line}
\end{table}

\boldmath
\section{Measurement of \boldmath$\Rb$}\label{s:rb}
\unboldmath

The fraction of \bbbar\ events within the hadronic event sample
$\Rb = \sigma_{\bbbar} / \sigma_{\rm{had}}$ is closely related to the ratio
of partial widths $\Rbz = \Gamma_{\bbbar} / \Ghad$ $(\Rbz = \Rb + 0.0002)$.
$\Rbz$ plays a special role within the electroweak quantities, since in the 
ratio of widths practically all corrections from the electroweak propagator  
and QCD effects cancel, so that $\Rbz$ only measures corrections to the 
Z\bbbar\ vertex.
At LEP $\Rbz$ is limited by systematic errors, mainly from three sources:
\begin{enumerate}
\item background from \ccbar\ events. 
This error arises mainly from physics
uncertainties in charm production and decay;
\item background from light quark events containing a gluon splitting
$g \rightarrow \bbbar$. 
This background is to a large extent irreducible and the error
is due to the uncertainty in the rate of the gluon splitting process;
\item hemisphere correlations from hard gluon radiation. Hard gluons
emitted at large angle take energy from both hemispheres. Since the
b-tagging efficiency is a function of the energy of the decaying
b hadron, this leads to a correlation between the hemispheres.
\end{enumerate}
The LEP analysis with the smallest error at present is from DELPHI \cite{drb}, 
and has about equal statistical and systematic uncertainties.
The DELPHI analysis uses a b-tag with an efficiency of 30\,\% and a 
purity of 98\,\%.
At TESLA a purity of 99.3\,\% at the same efficiency seems feasible
\cite{guennadi}, reducing
the statistical error by a factor of 20.
At the same time the charm background is reduced by a factor of four. Even 
assuming no improvements in 
the understanding of charm production and decay by the
time of TESLA, the systematics from this source will still be reduced 
accordingly.

The rate of gluon splitting into \ccbar\ and \bbbar\ pairs is measured at LEP. 
With the large event sample and  high quality b-tag there is no doubt 
that this can be measured much more precisely at TESLA.

Rejection of non \bbbar\ backgrounds is accomplished by exploiting the
long decay length and high invarient mass of b hadrons.
At TESLA, the impact parameter resolution (which determines the resolvablility
of secondary vertices)  is about a factor of 10 better than at
LEP, and the losses of \bbbar\ events will be mainly due to the invariant mass
cut. Since the invariant mass is a Lorenz invariant quantity the energy 
dependence of the b-tag should be much less than at LEP, so that for 
QCD hemisphere correlations a reduction of a factor four to five is also
realistic.

In summary a total error of $\Delta \Rb = 0.00014$ can be expected. This
is an improvement of a factor five compared to the present LEP average
\cite{lepew}.

\section{Measurement of \boldmath$\ALR$}\label{s:alr}

\subsection{The Blondel scheme for the linear collider}
In general the cross section $\ee \rightarrow \zb \rightarrow \ffbar$ for 
polarised beams can be written as
\begin{equation}
\sigma \, = \, \sigma_u \left[ 1 - \ppl \pmi + \ALR (\ppl - \pmi) \right],
\end{equation}
where $\ppl$ and $\pmi$ are the longitudinal polarisations of the positrons 
and electrons measured in the direction of the particle's velocity. $\sigma_u$
is the unpolarised cross section and $\ALR$ the left-right asymmetry.

With the assumption of an absolute electron/positron polarisation of
$\pmi = 80 \%$ and $\ppl = 60 \%$, and the possible polarisation states
$\pm \pmi$ and $\pm \ppl$, four statistically independent measurements are
possible:
\begin{eqnarray*}
\sigma_{++} & = & 
\sigma_u \left[ 1 - \ppl \pmi + \ALR (\pd \ppl - \pmi) \right] \\
\sigma_{-+} & = & 
\sigma_u \left[ 1 + \ppl \pmi + \ALR (- \ppl - \pmi) \right]\\
\sigma_{+-} & = & 
\sigma_u \left[ 1 + \ppl \pmi + \ALR (\pd \ppl + \pmi) \right] \\
\sigma_{--} & = & 
\sigma_u \left[ 1 - \ppl \pmi + \ALR (-\ppl + \pmi) \right] 
\end{eqnarray*}
where in $\sigma_{\pm\pm}$ the first sign denotes the positron and the second
one the electron polarisation.

The left-right asymmetry is then given by
\[
\ALR \, = \, \sqrt{\frac{
( \sigma_{++}+\sigma_{-+}-\sigma_{+-}-\sigma_{--})
(-\sigma_{++}+\sigma_{-+}-\sigma_{+-}+\sigma_{--})}{
( \sigma_{++}+\sigma_{-+}+\sigma_{+-}+\sigma_{--})
(-\sigma_{++}+\sigma_{-+}+\sigma_{+-}-\sigma_{--})}}
\]
without any need for polarisation measurements.

The statistical error is given as usual by 
\[
\Delta \ALR^2 \, = \, \sum_{i=\pm, j=\pm} 
\left( \frac{\partial\ALR}{\partial \sigma_{ij}} \Delta \sigma_{ij} \right)^2
\]
where the derivatives are given by
\begin{eqnarray*}
\frac{\partial\ALR}{\partial \sigma_{++}} & = & \frac{1}{8 \sigma_u \ppl\pmi}
\left[ -\ppl + \pmi - (\ppl \pmi -1) \ALR \right] \\
\frac{\partial\ALR}{\partial \sigma_{-+}} & = & \frac{1}{8 \sigma_u \ppl\pmi}
\left[ -\ppl - \pmi - (\ppl \pmi +1) \ALR \right] \\
\frac{\partial\ALR}{\partial \sigma_{+-}} & = & \frac{1}{8 \sigma_u \ppl\pmi}
\left[ \pd  \ppl + \pmi - (\ppl \pmi +1) \ALR \right] \\
\frac{\partial\ALR}{\partial \sigma_{--}} & = & \frac{1}{8 \sigma_u \ppl\pmi}
\left[  \pd \ppl - \pmi - (\ppl \pmi -1) \ALR \right] 
\end{eqnarray*}

Ignoring the terms proportional to $\ALR$ the derivatives for the large
cross sections $\sigma_{+-}$ and $\sigma_{-+}$ are proportional to 
$\ppl + \pmi$, whereas the ones at the small cross sections 
$\sigma_{++}$ and $\sigma_{--}$ are proportional to $|\ppl - \pmi|$.
In addition, for equal luminosity, the absolute errors on the large cross
sections are larger than on the small ones, so that the optimal distribution
of luminosities is around

\begin{equation}
\label{eq:lrat}
\frac{{\cal L}_{+-}+{\cal L}_{-+}}{{\cal L}} \, = \,
\sqrt{\frac{(1-\ppl\pmi)(\pmi-\ppl)^2}
{(1+\ppl\pmi)(\pmi+\ppl)^2+(1-\ppl\pmi)(\pmi-\ppl)^2}}
\end{equation}
Taking $\pmi=80\%$ and $\ppl=60\%$ the optimum is around 0.1, which means
that very little luminosity is needed for the small cross sections and
few \zb\ are lost. Figure \ref{f:opt} shows the statistical error on 
$\ALR$ for these polarisation values as a function of the luminosity spent on 
the small cross sections using $\ALR=0.15$. 
Since the minimum is very shallow no complicated optimisation procedure will 
be needed.

\epostscale{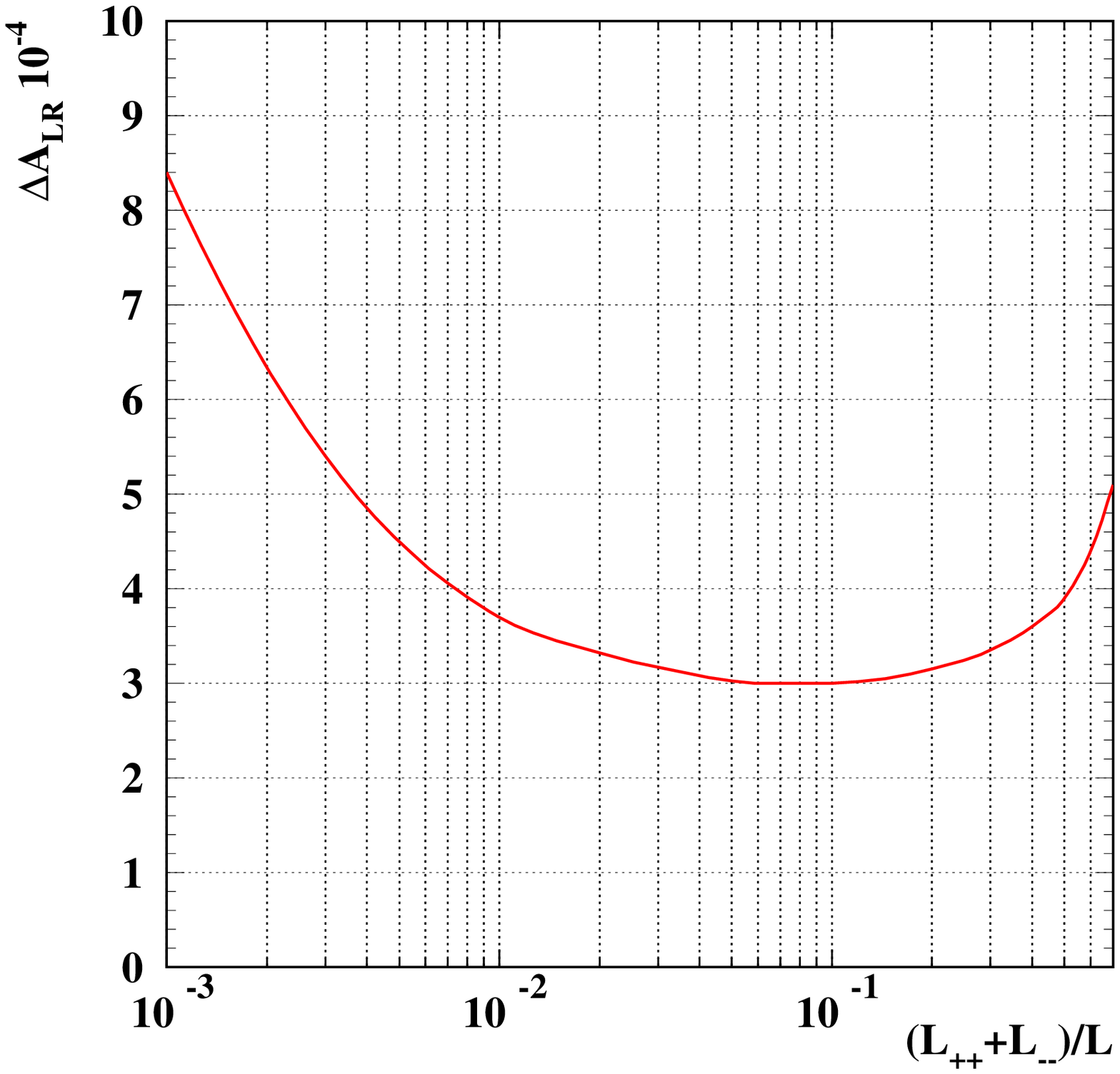}{f:opt}{0.7}{The statistical error on $\ALR$ as a 
function of the luminosity spent on the small cross sections.}

With $10^9$ recorded \zb\ an error of 
$\Delta \ALR = 3.0\cdot 10^{-5}$ can be reached. 
Figure \ref{f:dadpp} shows the statistical error as a function of the 
positron polarisation. For high values of $\ppl$ the error is nearly
constant, so that there is little to gain by increasing $\ppl$ much above
$60 \%$.

\epostscale{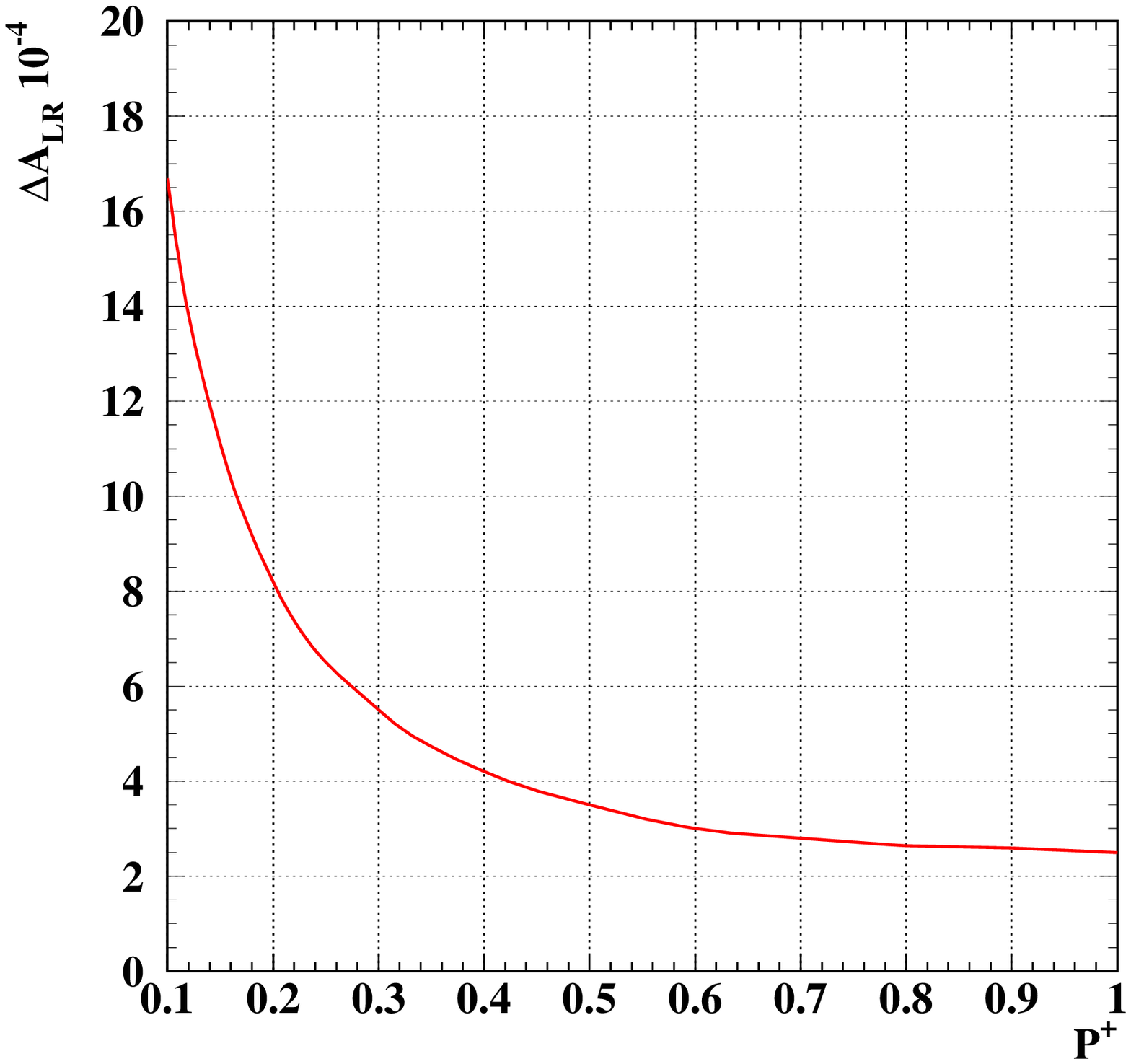}{f:dadpp}{0.7}{The statistical error on $\ALR$ as a 
function of the positron polarisation.}


Although no absolute polarisation measurements are needed in this scheme,
some relative measurements are still required. In particular the difference
between the absolute values of the positive and negative polarisations
of the same beam has a significant impact on the determination of $\ALR$. 
This will be treated in detail in section \ref{s:poldif}.

Other possible systematic errors should be well under control:
\begin{itemize}
\item The change of $\ALR$ with the centre of mass energy close to the peak
of the \zb-resonance is about $2 \cdot 10^{-2}/\rm GeV$, so 
the beam energy needs to be controlled to a precision of about 1\,MeV.
\item Non-resonant backgrounds need to be kept below $10^{-4}$. This
seems feasible from extrapolating the LEP experience.
\item To measure the luminosity differences between the different polarisation 
states a LEP-type luminometer measuring small angle Bhabha scattering is 
required. Again here only relative measurements need to be done.
\item  The amount of beamstrahlung expected in the high luminosity \zb\ 
running \cite{thorsten} changes $\ALR$ by
$\Delta \ALR \, = \, 9 \cdot 10^{-4}$. It thus needs to be known to a 
precision of a few percent.
\end{itemize}

\subsubsection{Measurement of the polarisation difference} \label{s:poldif}

In the scheme proposed here, a possible difference between the absolute values
of the positive and negative polarisations of the same beam can only be 
measured with polarimetry. It will thus be assumed that Compton polarimeters 
of the same type as that operated at SLD \cite{sldpol} are installed in the
electron and positron beams.

In this type of polarimeter the beam is scattered by a highly polarised laser
beam. The sign of the laser polarisation is chosen at random and the 
asymmetry of the scattered electrons is analysed by a detector 
(Cherenkov in the case of SLD). The possibility of varying the laser 
polarisation enables the two electron (positron) polarisation states 
to be measured independently. The polarisation measurement is vulnerable
to a possible polarisation dependent asymmetry of the laser currents.
This problem can be overcome if, as at SLD, the detector consists
of different channels with different analysing power.

With the simultaneous application of the Blondel scheme, the analysing powers
of the different detector channels need not be known {\rm a priori\/},
but can be obtained from data. The laser polarisation can be measured very 
precisely, and
enters only in a term  proportional to the small polarisation difference, 
so that its error is totally negligible.
Assuming analysing powers of 0.4 and 0 for the two detector channels the
error on $\ALR$ becomes
\[
\Delta \ALR \,= \, \Delta A_{\rm {LR}}^{(0)} \sqrt{1+8/x}
\]
where $\Delta A_{\rm {LR}}^{(0)}$ is the error on $\ALR$ without the 
contribution from the polarimeter and $x$ is the ratio of Compton events in 
each polarimeter channel to the number of \zb\ events. 
A value of $x \sim 10$, which seems no problem given the SLD experience, 
would thus increase the statistical error to 
$\Delta \ALR \, = \, 4 \cdot 10^{-5} \cdot \sqrt{\frac{10^9}{N_Z}}$.

\subsubsection{Alternative measurement of the polarisation difference}

If, in addition to the positive and negative polarisation, unpolarised 
beams are also used, nine different polarisation combinations are possible.
In this scheme the polarisation of the positive and negative states can be
measured separately from the data. The statistical error for this case 
increases to
$\Delta \ALR \, = \, 7 \cdot 10^{-5} \cdot \sqrt{\frac{10^9}{N_Z}}$.
However the residual polarisation of the unpolarised beam must be known to
better than $10^{-4}$ to get the systematic uncertainty from this source
below the statistical error.

\boldmath
\subsection{Interpretation of $\ALR$}
\unboldmath

Taking into account all the errors discussed above, a realistic estimate
for the final precision on $\ALR$ is $\Delta \ALR = 10^{-4}$.
This corresponds to an error of
0.000013 in $\swsqeffl$. At the Born level $\alpha,\, \GF$ and $\MZ$ enter
in the definition of $\sin^2 \theta$ and all give uncertainties on the 
prediction at least one order of magnitude smaller than the experimental
precision of the measurement.
However, at this level of precision, the effects of radiative corrections 
also become important, which is of course the main interest in doing the 
measurement. Therefore all the parameters of the Standard Model enter into
the prediction.

By far the largest correction comes from the running of the fine structure
constant to the \zb\ mass. Using only measurements of the hadronic cross 
section
at low energies, a value of $\alpha_{QED} (\MZ^2) = 1/(128.896 \pm 0.090)$
has been obtained~\cite{aem}, which translates into an error on
$\swsqeffl$ of 0.00023. By imposing QCD for energies above 1.7\,GeV the
uncertainty was recently reduced by about a factor of 
four~\cite{alpha_d,alpha_k,alpha_j}, but even this completely
dominates over the experimental error. 
If the hadronic cross section can be measured to a precision of
1\% for energies up to
about 3.6\,GeV, the error can be reduced by a total factor of nine. If
it is known to 1\% up to the $\Upsilon$ resonance at 9.5\,GeV,
the total reduction would be a factor
of 14 \cite{alpha_j}. This would correspond to $\Delta \swsqeffl = 0.000017$.

A top mass error of 1\,GeV would contribute an uncertainty of 0.000032 to
the $\swsqeffl$ prediction. At TESLA it should be possible to measure $\MT$
to 120\,MeV, so that the top contribution to $\Delta \swsqeffl$ will be
negligible.

A $5\%$ variation in the Higgs mass changes $\swsqeffl$ by 0.000025. If the
Higgs is found by the time the $\ALR$ measurement is done, its mass can
be measured with 200\,MeV precision, so that it does not contribute to
the uncertainty. If the Higgs is not found, the measurement of $\swsqeffl$ 
can be used to predict its mass to a precision of $\sim 5\%$.
To visualise this improvement, Figure \ref{f:higgs} shows the fit $\chi^2$
as a function of the Higgs mass for the present electroweak precision
data \cite{lepew} and for the TESLA expectation.

\epostfig{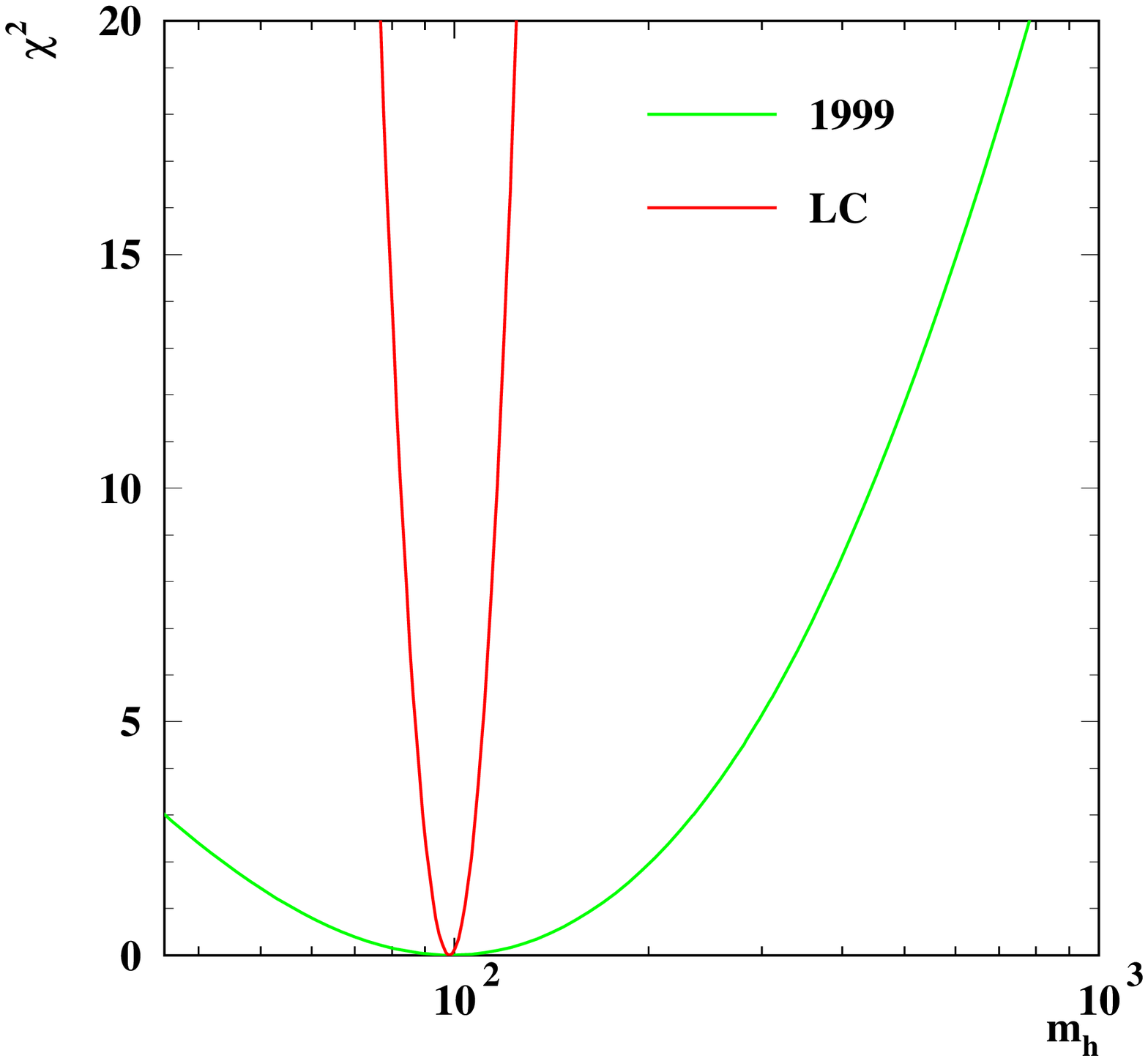}{f:higgs}{$\Delta \chi^2$ as a function of 
the Higgs mass for the electroweak
precision data now and after the linear collider \zb\ factory running.}

\section{Measurement of \boldmath$\cAb$}\label{s:ab}

At LEP the forward backward asymmetry for a quark $q$ measures the product of 
the coupling parameters for the initial state electron and the final
state quark:

\begin{eqnarray*}
A_{FB}^q & = & \frac{
  \sigma_{F}^{(q)}
- \sigma_{B}^{(q)}
}
{\sigma_{T}^{(q)}}\\
 & = & \frac{3}{4} \cAe \cAq
\end{eqnarray*}

If polarised beams are available the left-right-forward-backward asymmetry
is sensitive to the final state coupling only:

\begin{eqnarray*}
A_{FB,LR}^q & = & \frac{
  \sigma_{L,F}^{(q)}
- \sigma_{L,B}^{(q)}
- \sigma_{R,F}^{(q)}
+ \sigma_{R,B}^{(q)}
}
{\sigma_{L}^{(q)}+\sigma_{R}^{(q)}}\\
 & = & \frac{3}{4} {\cal P} \cAq
\end{eqnarray*}

Statistically this quantity is a a factor ${\cal P}/\cAe \sim 6$ more 
sensitive to $\cAq$ than $A_{FB}^q$. However most systematic errors are
proportional to the asymmetry, so are not reduced by using $A_{FB,LR}^q$ 
instead of $A_{FB}^q$.

At LEP two methods are used to measure $\Afbb$, leptons with 
high $p_t$ (transverse momentum with respect to the jet axis) and 
a jet charge measurement combined with a lifetime b-tag. With the better
b-tagging possibilities at a linear collider additional methods like
vertex-charge techniques are possible, as has been demonstrated at 
SLD \cite{sldab}. However in
the following only the extrapolation of the LEP methods will be discussed.

For the lepton measurement the ALEPH analysis \cite{aafbb} 
$(\Afbb = 0.1008 \pm 0.0043 (stat) \pm 0.0028 (syst))$ is taken as
a reference.
The statistical error with the assumed number of events will be
$\Delta \cAb = 4 \cdot 10^{-4}$. With this projected uncertainty the ALEPH
analysis would be completely dominated by the systematics from light and
especially charm quarks. However, this error can be reduced to the same
size as the statistical error using a b-tag with
$85 \%$ efficiency and a charm rejection of a factor 50, which seems
feasible \cite{guennadi}.

$\rm B\overline{B}$-mixing introduces a correction to the measured asymmetry
($\Afb^{meas} = \Afb(1-2 \chi)$) so that $\chi$ must be measured from double 
lepton events. This introduces an additional statistical error of
$\Delta \cAb = 9 \cdot 10^{-4}$. If the mixing analysis is done with the
same lepton selection as the asymmetry analysis, all uncertainties due to
the sample composition of the \bbbar\ events cancel.
Thus a total error of $\Delta \cAb = 11 \cdot 10^{-4}$ from the lepton analysis
should be possible.

For the jet charge method the recent DELPHI analysis \cite{dafbb}
$(\Afbb = 0.0982 \pm 0.0047 (stat) \pm 0.0016 (syst))$ will be extrapolated.
The statistical error is expected to become $\Delta \cAb = 5 \cdot 10^{-4}$.
With the same efficiency for the b-tag as in the DELPHI analysis (75\%) the
purity will be 99\% instead of 92\% \cite{guennadi} leading to a systematic
error from light quark backgrounds of $\Delta \cAb = 1 \cdot 10^{-3}$. Some
further optimisation in the efficiency-purity working point is clearly 
possible.

Another source of uncertainty is hemisphere correlations. With a 
cut on the thrust of the events (see fig \ref{f:thcut} \cite{britta})
the correlation can be brought down to the $10^{-3}$ level, so that even 
taking the full correlation as systematic uncertainty a total error of
$\Delta \cAb = 14 \cdot 10^{-4}$ from the jet charge analysis is possible.

\epostscale{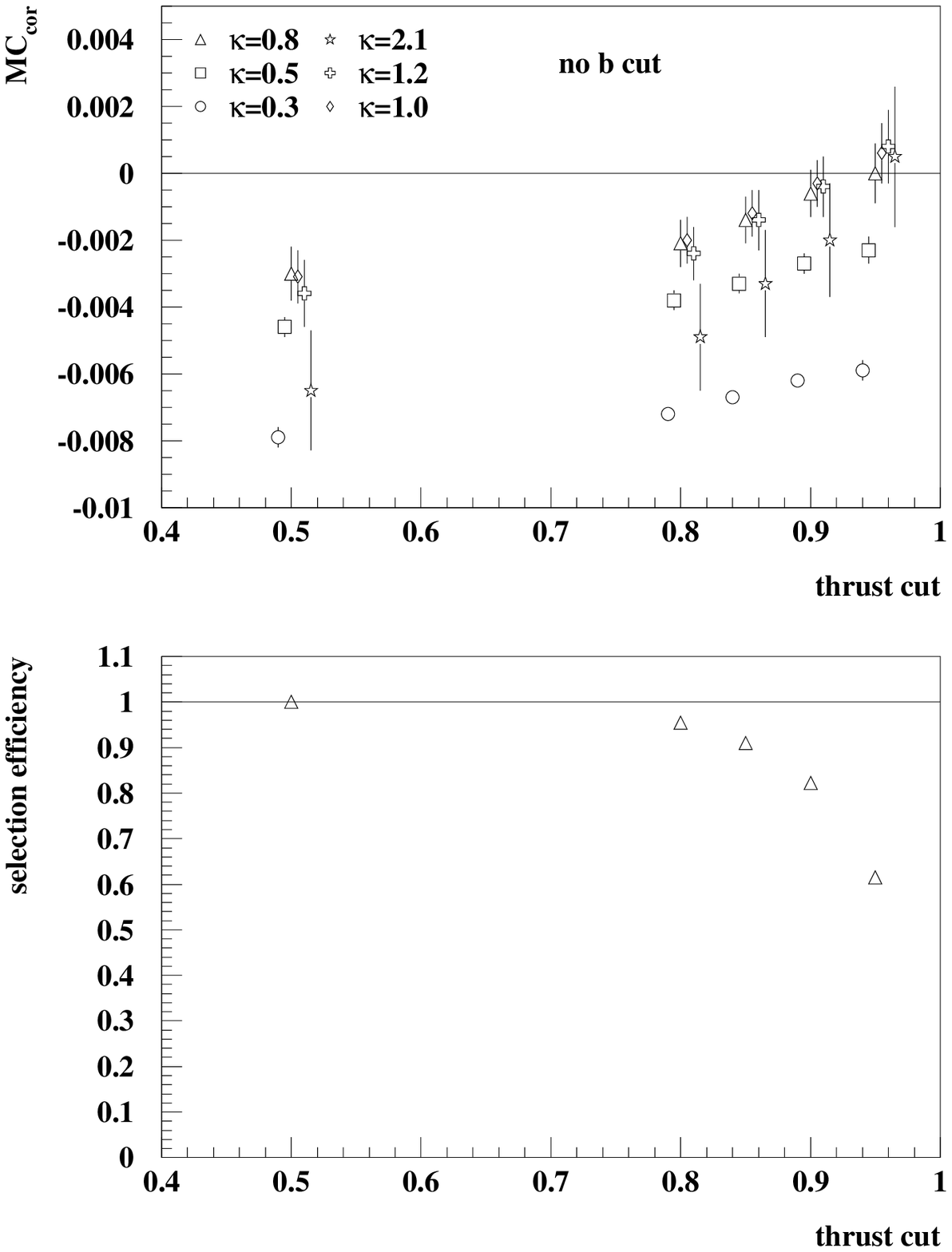}{f:thcut}{0.98}
{Hemisphere correlation for the $\Afbb$ jet charge analysis as a 
function of the thrust cut applied, taken from \protect\cite{britta}.}

Combining the two asymmetry analyses, 
a total error of $\Delta \cAb = 1 \cdot 10^{-3}$ 
seems realistic. As can be seen from Figure \ref{f:aeab}, this is an 
improvement of similar size as that obtainable on $\cAe$.

\epostscale{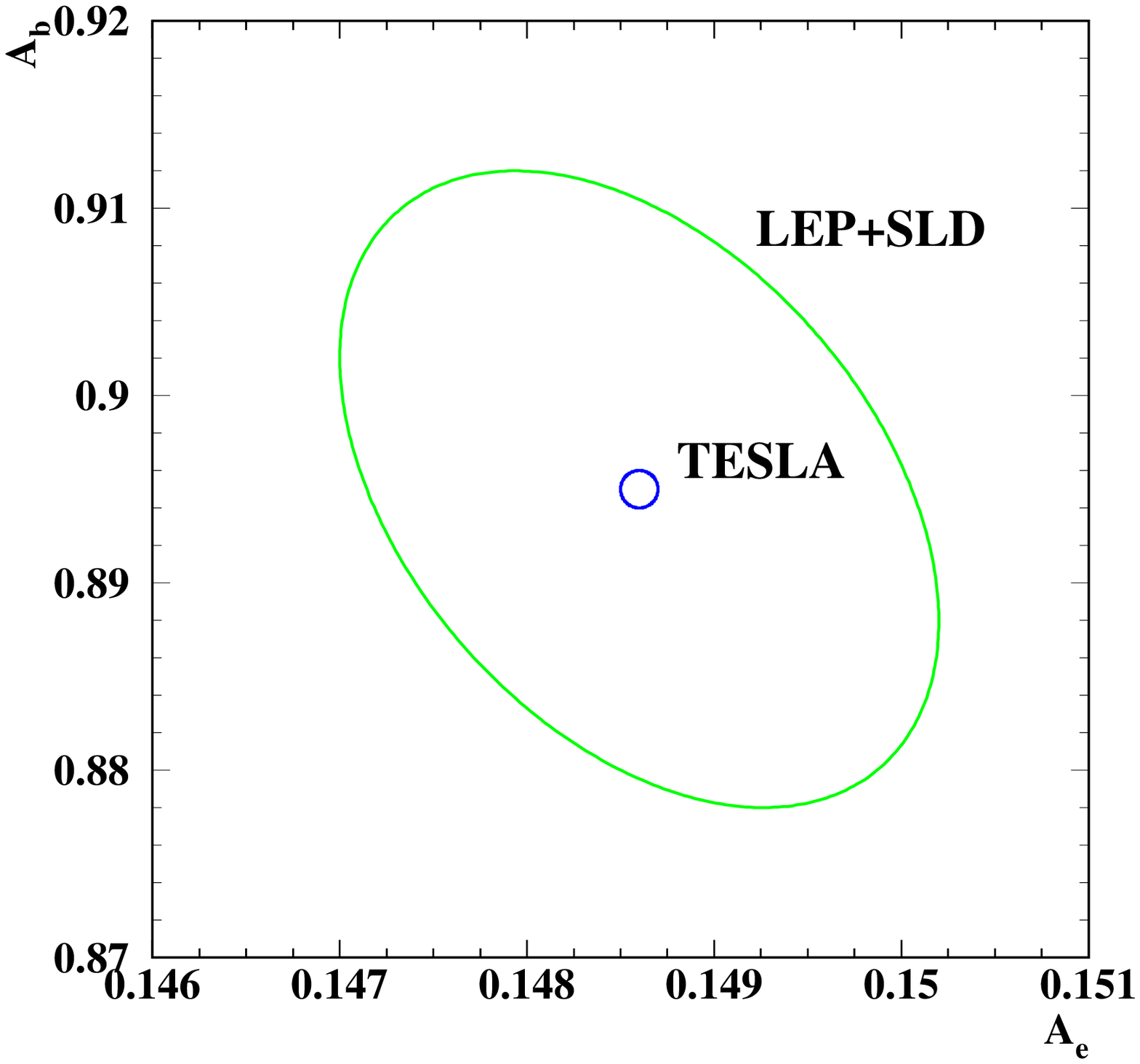}{f:aeab}{0.9}{$\cAb$ vs $\cAe$ from the
current LEP and SLD electroweak data,  compared to 
the expectation for the TESLA \zb\ factory running.}

\section{CP violation physics}\label{s:cp}

A data sample of $10^9$ hadronic \zb\ would include $4\times 10^8$ b hadrons,
which may allow competitive measurements of the angles of the CKM unitarity
triangle. The possibilities of measuring the angles $\alpha$, $\beta$,
$\gamma$ and
$\delta\gamma$ in both \bo\ and \bs\ decays is investigated in this
section, and the results compared with those achievable at BaBar, CDF, ATLAS
and LHC-b (taken as representative also of the results achievable at 
BELLE, D0, CMS and B-TeV respectively).

Some of the CP violation analyses make stringent demands on the performance
of the detector, and considerable advances over LEP detectors in the areas
of tracking and vertex reconstruction are required. Where necessary,
the `reference detector' described in the TESLA linear collider 
conceptual design report \cite{lccdr} is
assumed. It should be noted that no dedicated
particle identification devices ({\em e.g.} Cherenkov detectors) have been
added, but the use of ${\rm d}E/{\rm d}x$ information in the TPC is explored.
The detector performance has been studied using both
the fast simulation program SIMDET 3.1 \cite{simdet} and the full 
GEANT simulation BRAHMS 1.02 \cite{brahms}.
Both of these  simulation tools implement the detector geometry as described
in \cite{lccdr}.

\subsection{Initial state tagging}\label{s:itag}

Most of the CP violation studies of interest involve measuring the 
time-dependent asymmetry in the decay of \bo\ and \bobar\ mesons to a 
CP eigenstate $f_{\rm cp}$: $\bo,\bobar\rightarrow f_{\rm cp}$. The 
asymmetry $A(t)$ takes the form: 
\begin{equation}\label{e:asym}
A(t)\equiv\frac
{N_{\bo}(t)-N_{\bobar}(t)}{N_{\bo}(t)+N_{\bobar}(t)}
=\acs\cos\dm\,t + \asn\sin\dm\,t
\end{equation}
where \dm\ is the $\bo-\bobar$ mixing parameter. Since the final state is
a CP eigenstate ({\em e.g.\/} $\jpsi\ks$, $\pppm$), the initial state 
(\bo\ or \bobar) must be determined (`tagged') from information elsewhere
in the event. In most experiments,  this is done by exploiting the pair
production of \bbbar, and tagging the state (b or \bqbar) of the other quark
in the event (which in general does not decay to a CP eigenstate). However, 
both the efficiency and the purity of this procedure is low, leading to a 
significant `mistag' (events tagged incorrectly).

The use of polarised beams at TESLA, combined with the large forward-backward
asymmetry of the b quark coupling to the \zb, allows a different method to be
used, as at SLD \cite{sldtag}.
With polarised beams, the differential cross section for 
$\rm e^+e^-\rightarrow\zb\rightarrow\bbbar$ takes the form:
\begin{equation}
\frac{d\sigma}{d\cos\theta}=
(1-A_{\rm e}\pol)(1+\cos^2\theta)+2A_{\rm b}(A_{\rm e}-\pol)\cos\theta
\end{equation}
where $A_{\rm e}=0.15$ and $A_{\rm b}=0.93$ are the fermion couplings to the
\zb\,\cite{pdg98}, and $\pol=(\polem-\polep)/(1-\polem\polep)$ is the effective
polarisation. This cross section is plotted
in Figure~\ref{f:poltag}(a) for the expected polarisation levels of
$\polem=\pm 0.8$ and $\polep=\pm 0.6$. The very large asymmetry resulting
from the high level of effective polarisation when \polem\ and \polep\ have
opposite signs is clearly seen.

\epostfig{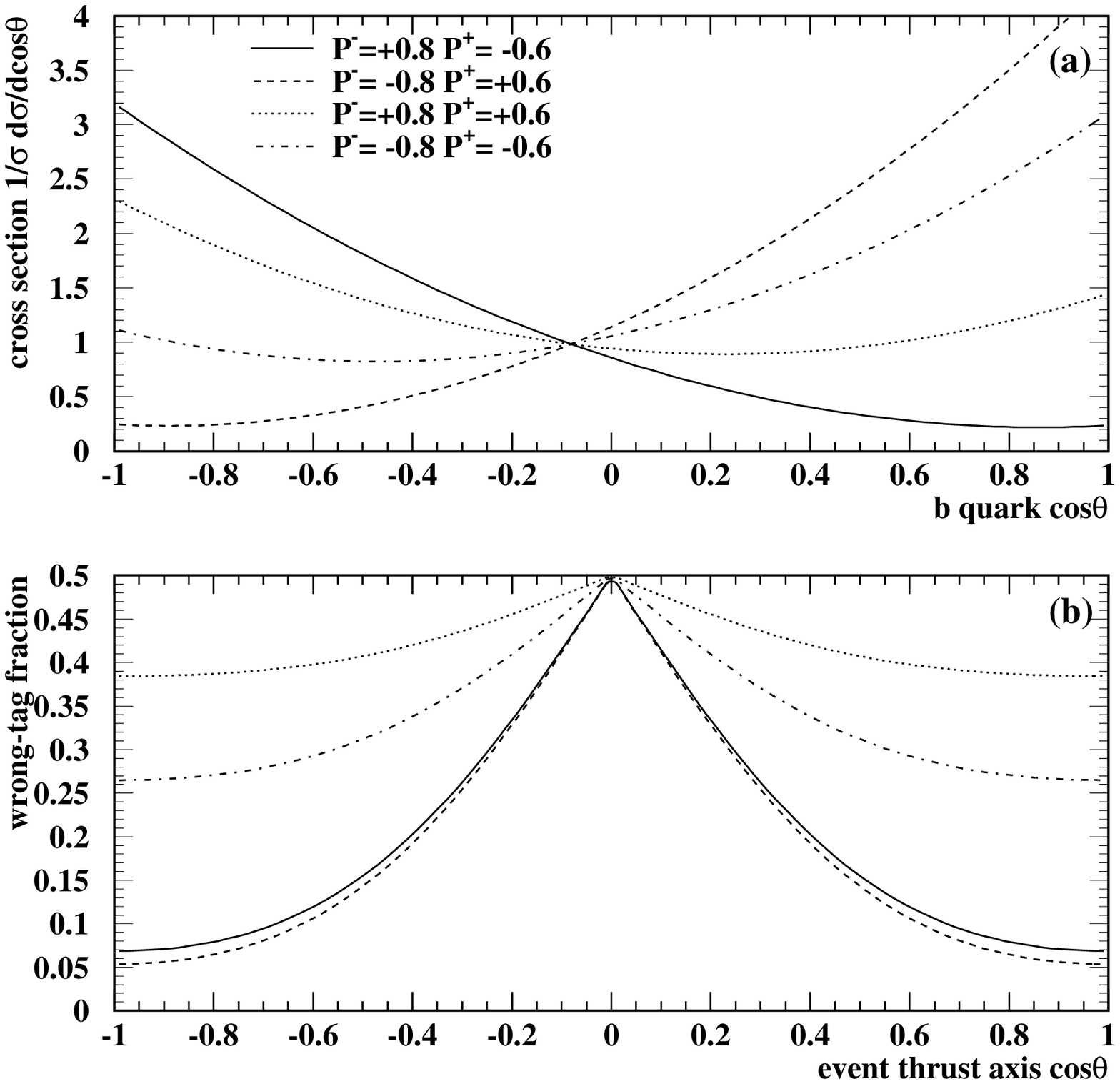}{f:poltag}{(a) Differential cross section for 
$\rm e^+e^-\rightarrow\zb\rightarrow\bbbar$ as a function of the b quark
polar angle $\cos\theta$, for various polarisations of the incoming 
electron and positron beams; (b) resulting wrong-tag fraction when the
b quark hemisphere is inferred from the thrust axis direction.}

This asymmetry can be used to tag the initial b quark flavour, by simply
assuming (for $\polem<0$) the forward ($\cos\theta>0$) hemisphere contains the b quark and the backward ($\cos\theta<0$) contains the \bqbar\ antiquark.
The resulting mistag rate $\eta$ 
is 0.5 (equivalent to purely random tag assignment)
when $\cos\theta=0$, but drops to {\em e.g.\/} only 6\,\% at 
$|\cos\theta|=0.9$. The tagging power can be quantified by means
of the `separation power' $s$, defined generally as $s=\epsilon D^2$ 
where $\epsilon$ is the tagging efficiency and $D$ the dilution 
$D=(1-2\eta)$ \cite{babarbook}. 
The dilution represents the reduction in the size of the
measured asymmetry due to the imperfect initial state tagging, and 
$s=\epsilon D^2$ gives the corresponding loss in the total number of 
events---one imperfectly tagged event is equivalent to $s$ perfectly tagged
events. With the polarisation tag, the mistag depends strongly on
$\cos\theta$, and the effective separation is given by the integral:
\[
<s>=\frac{\int_0^{\cos\theta_{\rm max}} \frac{d\sigma}{d\cos\theta}
(1-2\eta(\cos\theta))^2\,{\rm d}\cos\theta}
{\int_0^{\cos\theta_{\rm max}} \frac{d\sigma}{d\cos\theta}
\,{\rm d}\cos\theta}
\]
where $\eta(\cos\theta)$ is the mistag rate at any particular value of
$\cos\theta$. For $\polem=0.8$, $\polep=-0.6$ and $\cos\theta_{\rm max}=0.9$,
{\em i.e.\/} assuming a uniform event reconstruction efficiency up to
$\cos\theta=0.9$, the separation is $s=0.47$. For the 10\,\% of
the luminosity taken with same sign polarisation to minimise the error on
$\ALR$,
the separation is only $s=0.09$, and is not useful. These values should be
compared with typical values of $0.20$ in LEP \bo\  oscillation analyses, 
$0.06$ at CDF (for the $\sin 2\beta$ analysis \cite{cdfstb}), 
similar values of about
0.06 for future hadron collider experiments and 0.2--0.3 expected for
BaBar at an $e^+e^-$ b factory \cite{babarbook}.

It is clear that the initial state tagging with polarised beams offers
an important advantage over other facilities, and the tagging can 
be further improved by combining the polarisation information with
all the other tags used at LEP and SLD. A separation of $s=0.45$ has been
used throughout this note, and even higher values may be achievable with
appropriate combination of tags.

\subsection{Measurement of $\bf\beta$}\label{s:beta}

The cleanest way to measure the angle $\beta$ is to study the decay mode
$\bo,\bobar\rightarrow\jpsi\ks$, where the time dependent asymmetry
follows equation~\ref{e:asym} with $\asn=-\sin 2\beta$ and $\acs=0$.
The decay is easy to reconstruct with $\jpsi\rightarrow e^+e^-$ or
$\mu^+\mu^-$ and $\ks\rightarrow\pppm$, and first results have already
been obtained from existing experiments \cite{cdfstb,opalstb}. With 
$10^9$ \zb\ and a reconstruction efficiency of 30\,\% (OPAL achieved
20\,\% with inferior tracking and mass resolution), a total of
1900 reconstructed events would be expected. The effects of time resolution
and combinatorial background should be small, leading to an expected
error on $\sin 2\beta$ of $\pm 0.04$. This should be compared with
$0.12$ for BaBar in 1~year of running ($30\rm\,fb^{-1}$) \cite{babarbook},
$0.08$ for CDF from run II ($2\rm\,fb^{-1}$) \cite{mpaul}, $0.02$ from ATLAS in
1 year \cite{atlasp} and 0.01--0.02 from LHC-b \cite{lhcbtp}.

It is clear that in this relatively easy decay channel, the huge 
number of events reconstructed at LHC gives a decisive advantage. However,
$\sin 2\beta$ can also be measured in many other \bo\ decay modes, for example
$\jpsi\rm K^0_L$, $\jpsi\rm K^{*0}$, $\psi'\ks$, and $\rm D^{(*)}D^{(*)}$. 
An important part
of the b factories' physics program will be to measure $\sin 2\beta$ in all
these modes, and demonstrate consistent results. Some of these decay modes
are difficult at hadron colliders, but should be accessible in the cleaner
environment of the $e^+e^-$ linear collider. With the larger number of events 
and better tagging than at b factories, the linear collider should be able
to significantly improve on some of these measurements.

\subsection{Measurement of $\bf\alpha$}\label{s:alpha}

The measurement of $\alpha$ requires the study of a decay with a contribution
from $\rm b\rightarrow du\overline{u}$. The simplest such decay is
$\bo,\bobar\rightarrow\pppm$, whose asymmetry can be used to measure
$\sin 2\alpha$ in the same way as the asymmetry in 
$\bo,\bobar\rightarrow\jpsi\ks$ measures $\sin 2\beta$. However, in the
$\pppm$ decay, there are significant complications resulting from 
so called `penguin' diagrams, which destroy the simple relationship between
$\sin2\alpha$ and the measured asymmetry \cite{babarbook,lhcbtp}. In the
absence of penguin contributions, the asymmetry takes the form given by
equation~\ref{e:asym} with $\acs=0$ and $\asn=-\sin 2\alpha$, but penguin
contributions introduce decay amplitudes with different strong phases,
leading to $\acs\ne 0$ and $\asn\ne -\sin 2\alpha$. The measured values of
the coefficients
\acs\ and \asn\ can be used to extract $\sin 2\alpha$, providing the amount
of penguin contribution is known.

Unfortunately, the contribution of penguin decays is now expected to be
significant, as the penguin dominated decay $\bo\rightarrow\rm K^+\pi^-$ has 
been measured to have an unexpectedly large branching ratio of 
$(1.4\pm 0.3\pm 0.2)\times 10^{-5}$, whilst only an upper limit of 
$0.84\times 10^{-5}$ has been set on the branching ratio of 
$\bo\rightarrow \pppm$ \cite{cleorare}. This has two serious implications
for the determination of $\sin 2\alpha$ in the $\pppm$ channel:
\begin{enumerate}
\item The decay $\bo\rightarrow\pppm$ has a large background from
$\bo\rightarrow\rm K^+\pi^-$, which must be cleanly separated, since it is
also expected to have a (different) non-trivial CP violating asymmetry.
\item The relative contributions of tree and penguin decays in 
$\bo\rightarrow\pppm$ must be determined, to allow the extraction of
$\sin 2\alpha$ from the measured values of \acs\ and \asn. One promising
technique is to use an isospin analysis, which requires measurement
of the three branching ratios for the decays $\bo\rightarrow\pppm$, 
$\bo\rightarrow\pzpz$ and $\bplus\rightarrow\pi^+\pi^0$.
\end{enumerate}
Hence a determination of $\sin 2\alpha$ requires both a measurement of the
asymmetry in $\pppm$ and measurement of the three branching ratios
for the isospin analysis. These measurements are discussed in more detail
below.

\subsubsection{Asymmetry measurement for $\rm\bf\pi^+\pi^-$}

The reconstruction of $\bo\rightarrow\pppm$ decays has been studied with
the full detector simulation program BRAHMS, but simulating only the
tracking detectors. The candidate selection was based on the OPAL search
for this decay \cite{opalrare}. After requiring the direction of the
thrust axis to satisfy $|\cos\theta_{\rm T}|<0.9$, jets were reconstructed
using a cone algorithm with half-angle 0.65\,rad and minimum jet energy 
5\,GeV. All possible opposite-charge pairs of tracks in each jet
were then combined as possible \bo\ candidates, requiring the vertex
probability to exceed 1\,\%, the decay length significance $L/\sigma_L$ to
exceed 3 and the b hadron candidate energy to exceed 25\,GeV. Additionally,
the opening angle $\phi$ between the two tracks was required to be less
than 0.5 and the angle $\theta^*$ of the $\pi^+$ in the rest frame
of the \bo\ was required to satisfy $|cos\theta^*|<0.8$ \cite{opalrare}.
Finally the reconstructed mass of the candidate was required to be
in the range 5.258--5.299\,GeV.
The entire selection is 46\,\% efficient for $\bo\rightarrow\pppm$ and results
in about 900 selected events from $10^9$ \zb\ decays for a branching ratio of 
$1.2\times 10^{-5}$, as assumed in \cite{babarbook}.

\epostfig{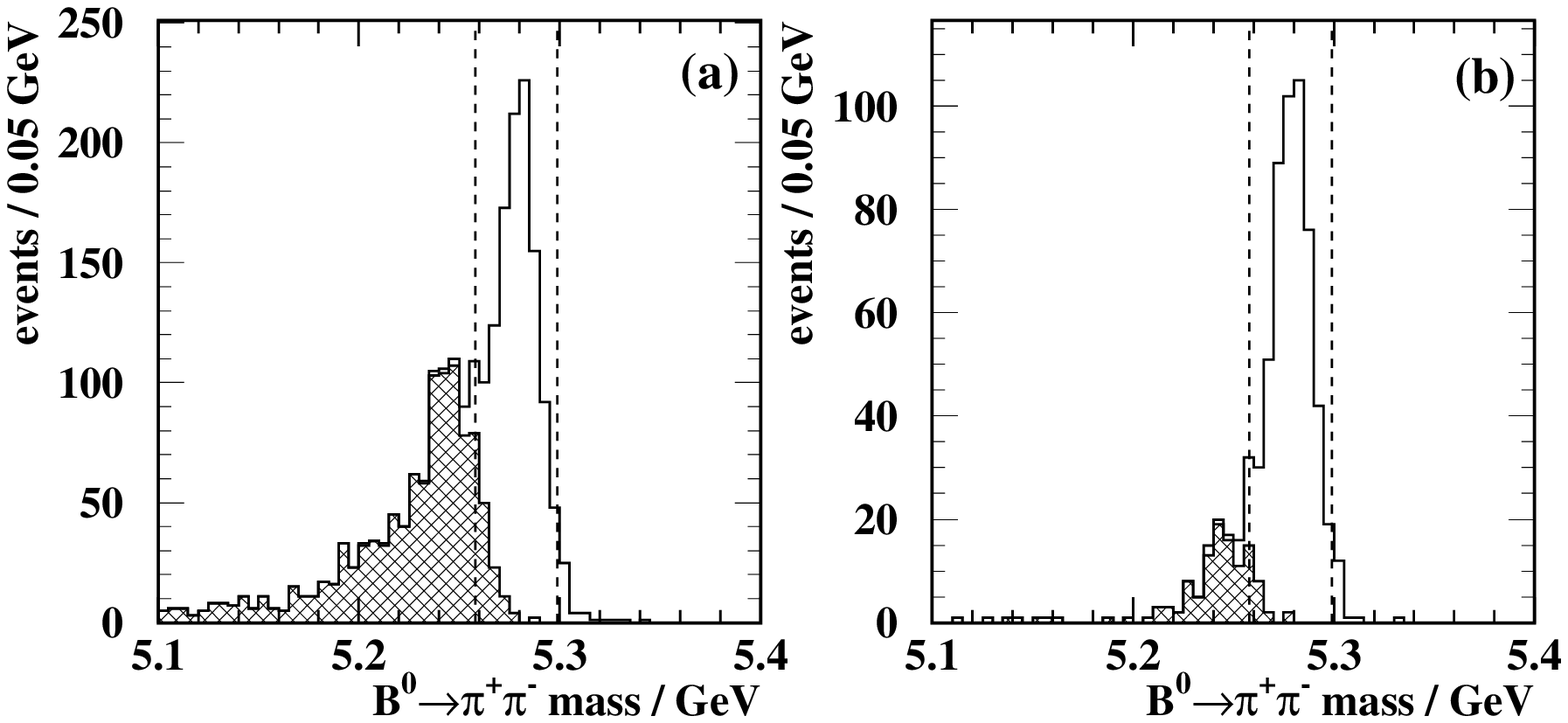}{f:b0sel}{Invariant mass distributions of 
 reconstructed $\bo\rightarrow\pppm$ decays (open histogram) and 
$\bo\rightarrow\rm K^+\pi^-$ decays (hatched histogram), without (a) and
with (b) a \dedx\ requirement. The positions of the mass cuts are shown by
the dotted lines.}

The resulting invariant mass spectrum for $\bo\rightarrow\pppm$ and 
$\rm K^+\pi^-$ decays is shown in Figure~\ref{f:b0sel}(a), assuming 
equal branching ratios for the two decay modes. The very good tracking
resolution of the linear collider detector gives a signal peak with RMS
of 11\,MeV, and this is enough to largely separate the signal from the
$\rm K^+\pi^-$ background. Within the mass region of 5.258--5.299\,GeV, the
signal to background ratio is about 10:1. This is in contrast to other
experiments, where particle ID has to be used to separate the two decays.
By making an additional cut of $\cos\theta^*>0$, exploiting the correlation
between the decay angle and the effective reconstructed mass for the
$K^+\pi^-$ decay, the background can be further reduced to a negligible level
for a 50\,\% loss in signal. 

The very good mass resolution allows the effective suppression of the
$\rm K^+\pi^-$ background without particle ID. However, the TPC of 
the proposed detector offers the possibility of also
using some \dedx\ information.
The performance of this has been estimated to give 1.5--2$\sigma$ separation
between pions and kaons in the momentum range of interest (10--30\,GeV)
\cite{hdedx}. This modest \dedx\ performance allows the signal to noise ratio
to be increased to 40:1 for a 50\,\% loss in signal efficiency, as shown
in Figure~\ref{f:b0sel}(b), before the application of the 
$\cos\theta^*>0$ cut. Although it is not needed in this particular channel,
the \dedx\ information may be useful in other studies ({\it e.g.\/} 3 body
decays) where the $\cos\theta^*$ cut cannot be applied.

Apart from the $\bo\rightarrow\rm K^+\pi^-$ decay, the main background
is expected to be combinatorial in nature.
This has been evaluated in two ways (i) with
a sample of $10^5$ \zb\ decays simulated in BRAHMS, and (ii) by making
extrapolations from the background seen in the OPAL search for the
$\bo\rightarrow\pppm$ decay \cite{opalrare}, accounting for the better
mass resolution of the linear collider detector. Both methods are statistically
limited, and lead to an expected  signal to noise ratio of 3:1 or better.

The precision to which the parameters \acs\ and \asn\ can be measured can be
used as a `figure of merit' for the various experiments. Assuming a 
signal to background level of 3:1, no residual contamination from 
$\bo\rightarrow\rm K^+\pi^-$ decays and no CP violation in the combinatorial
background, a simple fit to the asymmetry gives $\delta\acs=0.05$ and
$\delta\asn=0.07$ in $10^9$ \zb\ decays. 
The latter uncertainty can be interpreted as a statistical error on
$\sin 2\alpha$ in the absence of penguin contributions. The equivalent numbers
for other experiments are 0.26 for BaBar \cite{babarbook},
0.10 for CDF \cite{mpaul}, 0.14 for ATLAS \cite{atlasp} and 0.05 for LHC-b
\cite{lhcbtp}, from the same integrated luminosities as discussed in 
section~\ref{s:beta}. It should be noted that ATLAS has no particle ID
and insufficient mass resolution to separate the $\pppm$ and $\rm K^+\pi^-$
final states, so the analysis will be very difficult in the presence of 
this background. It can be seen that in this measurement the linear collider
is very competitive with other experiments, offering comparable precision
to LHC-b in a very different environment.

\subsubsection{Branching ratio measurements for $\rm\bf\pi^0\pi^0$ and $\rm\bf\pi^+\pi^0$}

The extraction of the penguin and tree amplitudes in $\bo\rightarrow\pppm$
can be achieved by measurement of the three branching ratios 
$\bo\rightarrow\pppm$, $\bo\rightarrow\pzpz$ and $\bplus\rightarrow\pi^+\pi^0$.
The first can be easily be measured as a by product of the asymmetry analysis
discussed above, but the other two are more difficult. In particular,
\bratio{\bo}{\pzpz}\ is expected to be extremely small, theoretical predictions
being of the order $10^{-6}$ and the experimental upper limit $9\times 10^{-6}$
\cite{cleorare}.

The feasibility of measuring the latter two branching ratios has been
studied using the fast simulation SIMDET, which contains a parameterisation
of the calorimeter response to overlapping particles derived from the
full BRAHMS simulation. The reconstruction is similar to that used for
the \pppm\ channel, but requires the explicit reconstruction of
$\pi^0$ mesons in the calorimeter. The $\pi^0$ have a mean energy of
about 15\,GeV, at which the two photons from the $\pi^0$ decay have 
a typical angular separation of 0.04\,rad, corresponding to
a spatial separation of 5\,cm at the calorimeter front face. Since
the granularity of the proposed electromagnetic calorimeter is 
0.015\,rad \cite{lccdr}, 
many $\pi^0$ are not clearly resolved into 2~photons but
appear as a single wide electromagnetic cluster. Both pairs of 
identified photons
with an invariant mass between 0.1 and 0.2\,GeV, and single
identified photons were therefore considered as $\pi^0$ candidates.
The photon identification was taken directly from SIMDET, which includes
an `energy flow' matching algorithm to identify neutral and charged
particles from the reconstructed tracks and calorimeter clusters. 

Reconstructed $\pi^0$ candidates were combined in pairs or with charged
hadrons to form \bo\ and \bplus\ candidates, in both cases requiring
the energy of each pion to exceed 8\,GeV and that of the b~hadron
to exceed 25\,GeV. The resulting invariant mass spectra in signal
events containing $\bo\rightarrow\pzpz$ or $\bplus\rightarrow\pi^+\pi^0$
are shown in Figure~\ref{f:b0pi0}. The invariant mass resolutions
are dominated by the $\pi^0$ reconstruction, and are 160 and 120\,MeV 
respectively, much worse than in the $\pppm$ mode. This resolution
is dominated by angular rather than energy resolution effects.
Large tails to lower masses are also visible in the reconstructed
mass distributions, caused by $\pi^0$ overlapping with other particles (from
fragmentation) in the same jet, spoiling the energy measurement.

\epostfig{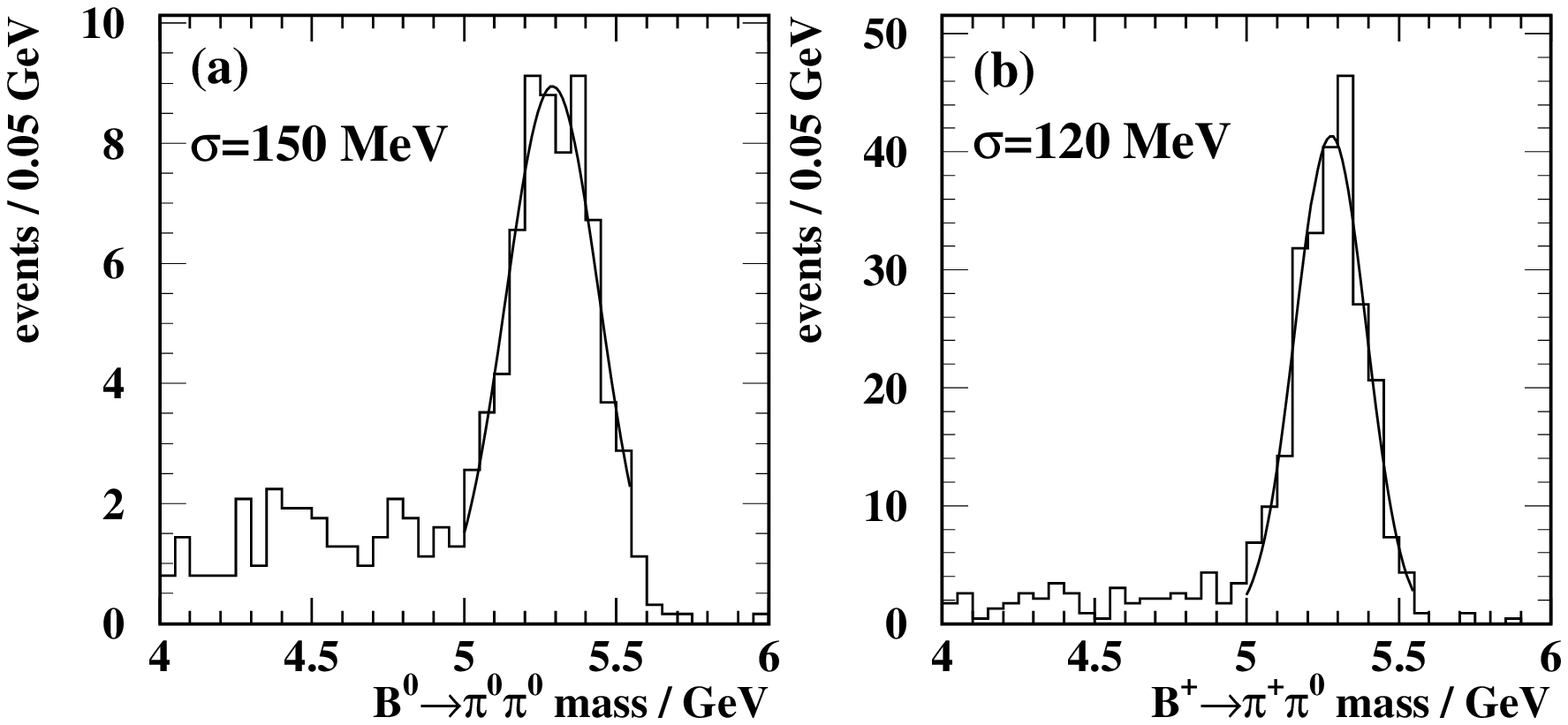}{f:b0pi0}{Reconstructed invariant mass distributions
for (a) $\bo\rightarrow\pzpz$ and (b) $\bplus\rightarrow\pi^+\pi^0$
signal events. The mass resolutions (excluding the tails of low mass events)
are indicated.}

Due to the lower mass resolution and difficult $\pi^0$ reconstruction,
these channels suffer from very large combinatorial background. This can
be reduced by applying a b-tag on the opposite hemisphere of the 
event---an efficiency of 50\,\% for \bbbar\ events with negligible 
contribution from other flavours was assumed. The remaining combinatorial
background was assessed using a sample of $2\times 10^6$ \bbbar\ events
processed through SIMDET. These studies show that the background 
is composed mainly of random tracks and clusters from b decays and 
fragmentation forming the correct invariant mass. It can be reduced by
applying a track based anti b-tag in the signal hemisphere (after first 
removing the candidate $\pi^+$ track in the $\pppz$ case). The
number of charged tracks from b decays in selected background hemispheres
is shown in Figure~\ref{f:b0pi0bg}(a) and~(b). Although the b decay 
multiplicity is lower than average (about 5) in these hemispheres, a veto
on any tracks with significant impact parameter can still 
be used to effectively suppress this background. A rejection factor
of 10 ({\em i.e.\/} 90\,\% of background events rejected) for negligible
loss of signal has been assumed.

\epostfig{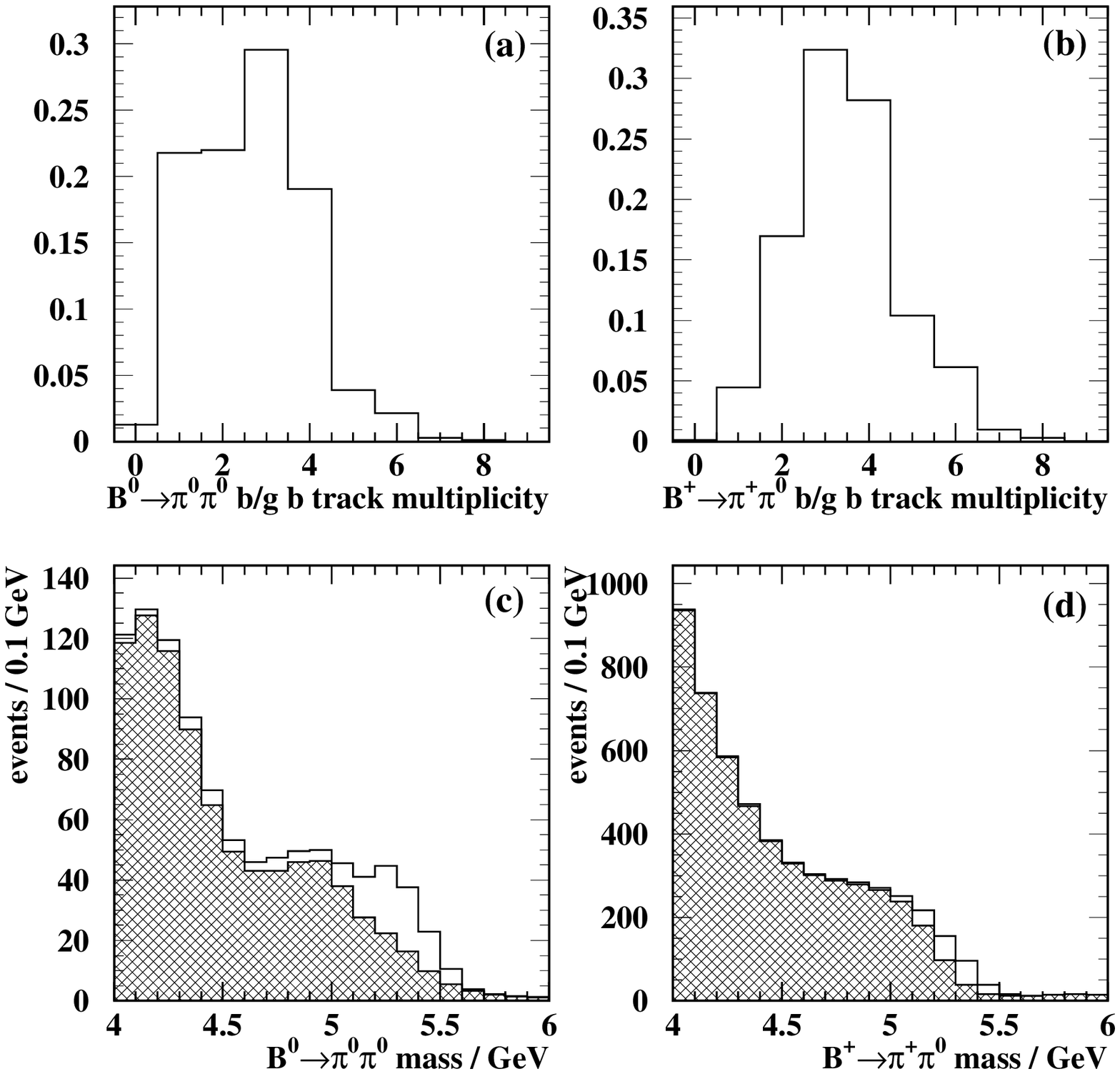}{f:b0pi0bg}{Background to $\bo\rightarrow\pzpz$ and
$\bplus\rightarrow\pppz$ selections: Number of tracks from 
b decays in (a) $\pzpz$ and (b) $\pppz$ background candidate hemispheres 
passing all cuts except signal hemisphere b-tag requirement; resulting
mass spectra for signal and background events for (c) $\pzpz$ and
(d) $\pppz$ decays. The signal is shown by the open histogram and
the background by the hatched histogram in each case. Signal 
branching ratios of $\bratio{\bo}{\pzpz}=3\times 10^{-6}$ and 
$\bratio{\bplus}{\pppz}=5\times 10^{-6}$ have been assumed.}

After applying the anti-b tag in the signal hemisphere, the expected numbers
of signal and background events in the mass window 5--5.5\,GeV
for various different branching ratio
assumptions are listed in Table~\ref{t:b0pi0}. The expected statistical errors
on the branching ratio, taking into account the numbers of signal and 
background events and no uncertainty on the background level, are also given,
and compared to those expected from BaBar after 10 years of running at
nominal luminosity ($300\rm\,fb^{-1}$) \cite{babarbook}. It can be seen
that these errors are somewhat larger than those from BaBar, and are 
in fact similar to those expected after only 5~years of BaBar running. 

\begin{table}
\centering 

\begin{tabular}{cc|cc|cc}\hline
Decay & $\rm {\cal B} /10^{-6}$ & Signal & Background & 
\multicolumn{2}{c}{$\rm \sigma_{\cal B}/\cal B$} \\
 & & events & events & LC & BaBar \\
\hline
$\bo\rightarrow\pzpz$ & 1 &  26 & 107 & 0.44 & 0.29 \\
                      & 2 &  52 & 107 & 0.24 & 0.17 \\
                      & 3 &  78 & 107 & 0.17 &  -   \\
                      & 5 & 130 & 107 & 0.12 & 0.08 \\ \hline
$\bplus\rightarrow\pppz$ & 2.5 &  95 & 577 & 0.27 & 0.20 \\
                         & 5   & 190 & 577 & 0.15 & 0.11 \\
                         & 7.5 & 285 & 577 & 0.10 & 0.08 \\
                         & 10  & 380 & 577 & 0.08 & 0.06 \\
\hline
\end{tabular}
\caption{\label{t:b0pi0}Number of signal and background events, and
resulting fractional statistical error $\sigma_{\cal B}/{\cal B}$ on
the branching ratio, for various values of $\bratio{\bo}{\pzpz}$ and
$\bratio{\bplus}{\pppz}$. Results for BaBar, assuming $300\,\rm fb^{-1}$
integrated luminosity, are also given \cite{babarbook}.}
\end{table}

The expected invariant mass spectra for $\bratio{\bo}{\pzpz}=3\times 10^{-6}$
and $\bratio{\bplus}{\pppz}=5\times 10^{-6}$ are shown in 
Figure~\ref{f:b0pi0bg}(c) and~(d). The background histograms are extrapolations
from the small number of surviving \bbbar\ events, so their shapes are 
somewhat uncertain. Nevertheless, it is clear that a good understanding
of the background shape will be necessary to measure the signal branching 
ratios, and this may be rather difficult to achieve.

In summary, the measurement of the branching ratios $\bratio{\bo}{\pzpz}$
and $\bratio{\bplus}{\pppz}$ would be an important contribution to the
understanding of CP violation in b decays, since it cannot be performed
at hadron colliders. However, it looks rather difficult with the present
detector and uncertainties in the level of background. To proceed further 
probably requires a more realistic calorimeter simulation and a more
detailed study of the $\pi^0$ reconstruction and background rejection.

\subsection{Measurement of $\rm\bf\Delta m_s$,  $\bf\gamma$ and $\bf\delta\gamma$}

The study of time dependent asymmetries in \bs\ decays provides a method
of determining the CKM angle $\gamma$ and the small angle $\delta\gamma$ 
(which is the `tilt' of the second unitarity triangle formed from
the relation 
$V^*_{\rm ud}V_{\rm td}+V^*_{\rm us}V_{\rm ts}+V^*_{\rm ub}V_{\rm tb}=0$).
\bs\ mesons are not produced at all in $\Upsilon(4S)$ decays (so are 
inaccessible to $e^+e^-$ b factories unless they run at a higher resonance)
and are only produced in 10\,\% of b decays at higher energies. The 
\bs\ mixing frequency \dms\ is also at least 25 times higher
than that for \bo\ mesons, so requiring very good time resolution to see
the oscillations and asymmetries. Both of these factors make studies of 
\bs\ decays much more challenging experimentally than studies of \bo\
decays.

\subsubsection{Measurements in $\rm\bf B_s\rightarrow D_s^-K^+$ and 
$\rm\bf B_s\rightarrow D_s^-\pi^+$}

Searches for \bs\ oscillations at LEP, SLD and the Tevatron have 
so far been carried out mainly using semi-inclusive modes such as
$\bs\rightarrow\rm D_s\ell\nu X$ and $\bs\rightarrow\rm \ell\nu X$. In these
analyses, the energy of the \bs\ (which is necessary to calculate the
decay proper time) is estimated using reconstructed and missing energy.
However, the limit on \dms\ is now $12.4\rm\,ps^{-1}$ \cite{dmslim}, and
measuring significantly higher frequencies can only be done with rare
fully reconstructed decays. The decay chain $\bs\rightarrow\rm D_s^-\pi^+$,
with $\rm D_s^-\rightarrow \phi\pi^-$, $\phi\rightarrow\rm  K^+K^-$ provides
a suitable decay mode to measure or set a limit on \dms, whilst the
similar $\bs\rightarrow\rm D_s^\mp K^\pm$ provides a way of measuring
$\gamma-2\delta\gamma$.

These channels have been studied using the full GEANT simulation BRAHMS.
The decay chain was fully reconstructed, requiring two oppositely charged
tracks to make a $\phi$ candidate with $1.015<m_\phi<1.025$\,GeV, combined
with another track to make a \ds\ candidate with $1.959<m_{\rm D_s}<1.979$\,GeV
and a further opposite signed track to make a \bs\ candidate. The 
3 tracks from the \ds\ were required to make a good vertex, as were
the extrapolated track from the \ds\ and the $\pi^+$ from the \bs. The latter
vertex was used to define the \bs\ decay length. The \bs\ energy was 
additionally required to exceed 20\,GeV, but no  particle ID (\dedx) cuts were
applied. The resulting mass distribution is shown in Figure~\ref{f:bsrec}(a),
for $\bs\rightarrow\rm D_s^-\pi^+$ signal events, and the same events 
reconstructed assuming the $\pi^+$ from the \bs\ decay to have the kaon mass.
The latter distribution, which forms a `reflection' background for the 
reconstruction of
$\bs\rightarrow\rm D_s^\mp  K^\pm$ is reasonably separated from the signal,
without the need for particle ID. The signal mass resolution is about 8\,MeV,
without the application of any kinematic fit.

\epostfig{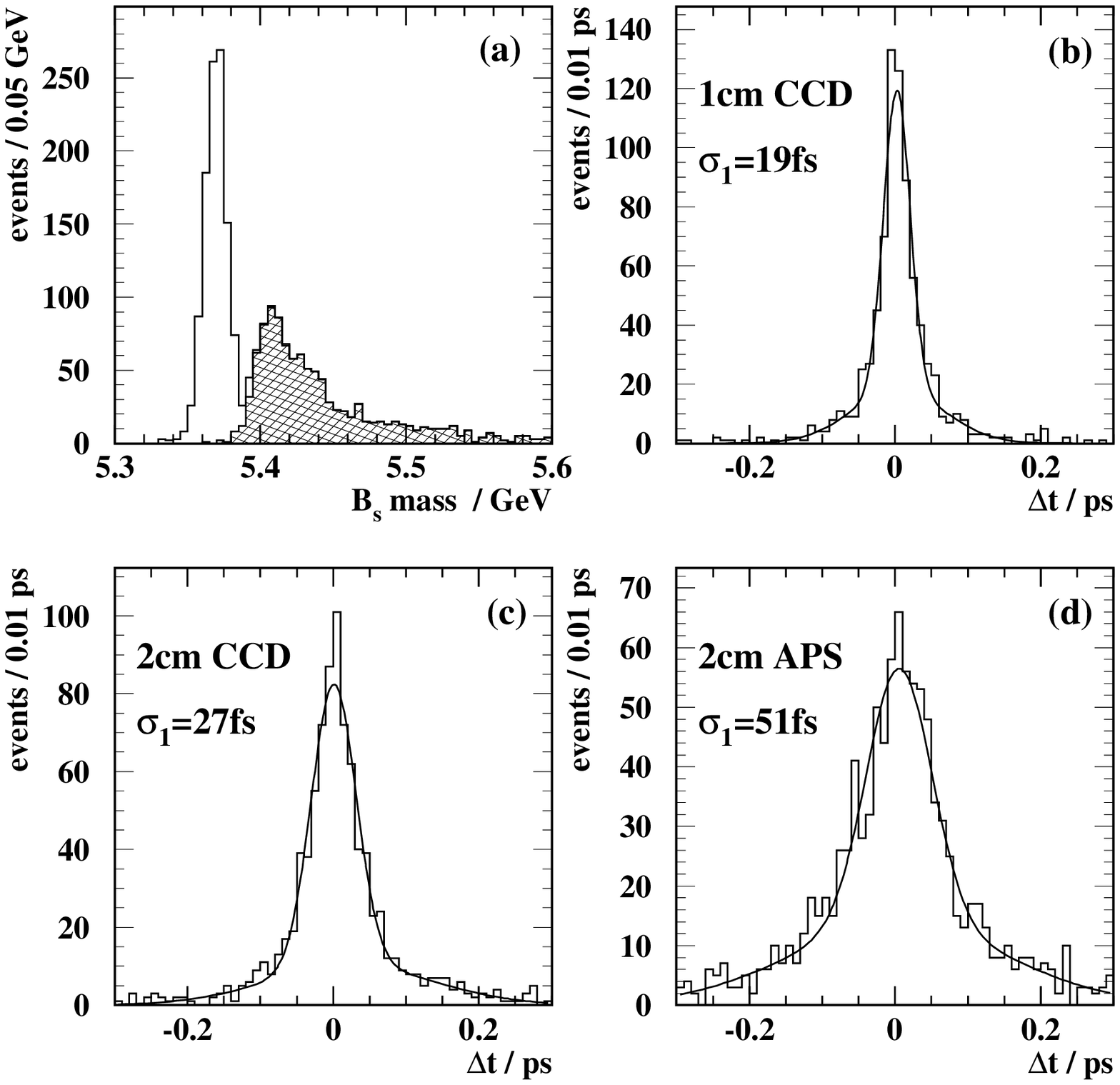}{f:bsrec}{(a) Reconstructed invariant mass distributions
for signal $\bs\rightarrow\rm D_s^-\pi^+$ (open histogram) and the
same events reconstructed as $\bs\rightarrow\rm D_s^-K^+$ (hatched
histogram); (b)--(d) 
reconstructed proper time resolutions for $\bs\rightarrow\rm D_s^-\pi^+$
events with different vertex detector options. Double Gaussian fits to the
resolutions are superimposed, and the widths of the narrow Gaussians 
($\sigma_1$) are indicated.}

The reconstruction efficiency for this selection is about 20\,\%, but many of
the losses occur in requiring good vertices, which may be improved with better
tracking code. Assuming $\bratio{\bs}{\rm D_s^-\pi^+}=3\times 10^{-3}$ 
(by analogy with $\bo\rightarrow\rm D^-\pi^+$ \cite{pdg98}) and 
$\bratio{\rm D_s^-}{\phi\pi^-}=3.5\,\%$ \cite{pdg98} gives a
total of about 500 reconstructed events in $10^9$ \zb\ decays.

The proper time of each decay was reconstructed from the measured decay
length (between the reconstructed primary vertex and the intersection
of the $\rm D_s^-$ and $\pi^+$ tracks) and reconstructed energy of each 
candidate. The energy resolution is extremely good (0.5\,\%) and the 
proper time resolution is dominated by the decay length reconstruction.
This was simulated separately for each of the vertex detector options 
available in BRAHMS---CCD vertex detectors surrounding  1\,cm and 2\,cm 
radius beampipes, and an APS vertex detector with a 2\,cm beampipe.
The resulting proper time resolutions are shown in 
Figure~\ref{f:bsrec}(b)--(d), together with the resolutions obtained from
a double Gaussian fit to the distribution of reconstructed minus true
proper time. The resolution clearly depends strongly on which vertex detector
option is chosen.

The sensitivity to \bs\ oscillations as a function of \dms\ can be quantified
by means of the amplitude method \cite{nimosc}. The amplitude $A$ of 
oscillations at each `test' value of \dms\ is measured, in a method
analogous to Fourier analysis. Oscillations are excluded if the value $A=1$
is excluded by the measurement, whilst exclusion of $A=0$ is required to
establish a signal. The significance of the analysis is then quantified
by the inverse of the error on $A$, $\sigma_A^{-1}$, which must be at least
5 to claim a signal (`$5\sigma$ discovery'), and at least $1.645$ to 
set a 95\.\% confidence level exclusion. The expected error can be calculated
approximately by:
\[
\sigma_A^{-1} = \sqrt{\frac{N}{2}}\fsig (1-2\eta) \exp (-\frac{1}{2}(\sigma_t\dms)^2)
\]
where $N$ is the number of events in the sample, \fsig\ the signal purity,
$\eta$ the mistag and $\sigma_t$ the proper time resolution. The results
of this calculation for $\fsig=0.5$ and $\fsig=0.9$ are shown in 
Figure~\ref{f:dmslim} for each of the vertex detector options.
The background to the $\bs\rightarrow\rm D_s^-\pi^+$ signal is expected to
be combinatorial and very small, due to the very good mass resolution and
two intermediate mass constraints ($\phi$ and \ds). Hence values of
\fsig\ close to one are expected to be realistic, although this has not
been studied with simulation. With $\fsig=0.9$, it can be seen that
\bs\ oscillations can be discovered between 20 and $40\rm\,ps^{-1}$, 
depending on the vertex detector options. This is comparable with the 
reach of CDF and ATLAS, though slightly lower than the $50\rm\,ps^{-1}$ 
expected by LHC-b \cite{lhcbtp}. The larger reach of LHC-b is due 
to the much higher numbers of reconstructed events---30000 as opposed to
500 for the linear collider, though the proper time resolutions are similar
and the tagging better at the linear collider.
However, such high values of \dms\ should also be visible through
their effects on the \bs\ width difference ($\Delta\Gamma/\Gamma$), and
are currently disfavoured within the framework of the Standard Model.

\epostfig{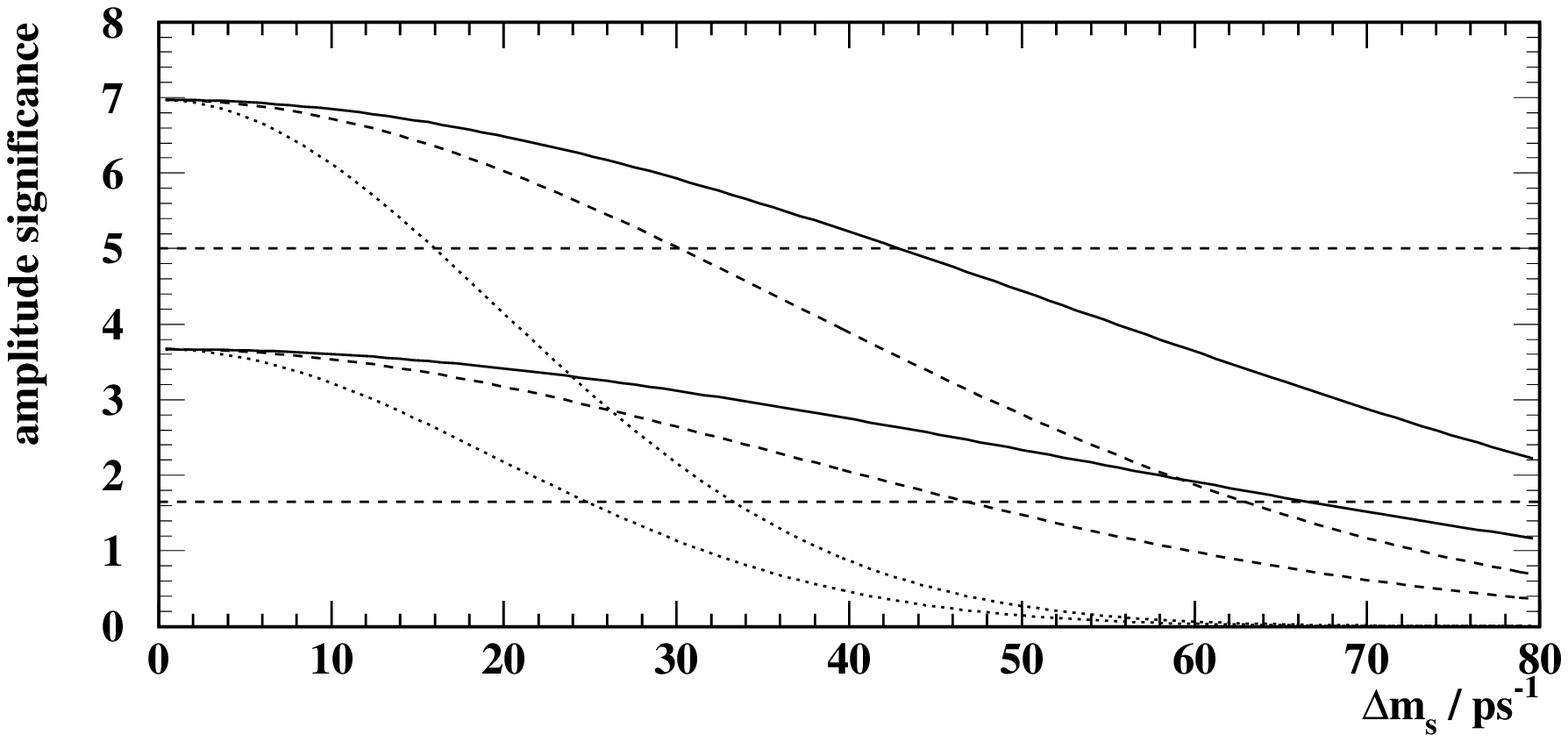}{f:dmslim}{Amplitude significance $\sigma_A^{-1}$
vs \dms\
curves for $\bs\rightarrow\rm D_s^-\pi^+$ oscillation analyses assuming a 
CCD vertex detector with 1\,cm radius beampipe (solid line), CCD vertex 
detector with 2\,cm beampipe (dashed line) and APS vertex detector with 2\,cm 
beampipe (dotted line). The upper set of curves refer to a signal fraction
of 90\,\%, the lower to a signal fraction of 50\,\%. The $5\sigma$ discovery
and $1.65\sigma$ exclusion significances are indicated.}

The decay $\bs\rightarrow\rm D_s^\mp K^\pm$ can be used to measure 
the CKM angle $\gamma-2\delta\gamma$. 
However, its branching ratio is expected to be at
least an order of magnitude smaller than that for 
$\bs\rightarrow\rm D_s^-\pi^+$,
so only around 50 reconstructed events are expected in $10^9$ \zb\ 
decays---not enough to measure the 4 time dependent decay rates necessary
for this analysis. LHC-b expects around 2500 events per year in this channel
\cite{lhcbtp} so the linear collider is clearly not competitive, even with
the large statistical advantage in tagging power.

\subsubsection{Measurements in $\rm\bf B_s\rightarrow \jpsi\phi$}

The decay $\bs\rightarrow\jpsi\phi$ is the \bs\ analogue of the
$\bo\rightarrow\jpsi\ks$ decay, and can be used to measure
 the CKM angle $\delta\gamma$. This asymmetry is expected to be very small
in the Standard Model, so a significant asymmetry would be a sign of
new physics. However, as with the other \bs\ decay modes discussed above,
the relatively small number of \bs\ mesons produced at the linear collider
make it uncompetitive with LHC-b, and also with 
ATLAS and CMS in this decay channel which is easy to trigger on.

The reconstruction of $\bs\rightarrow\jpsi\phi$ is similar to 
$\bo\rightarrow\jpsi\ks$, and with a 30\,\% reconstruction efficiency
700 events are expected in $10^9$ \zb. ATLAS and LHC-b expect around
40000 events \cite{atlasp,lhcbtp}, so again the linear collider is at 
a severe statistical disadvantage.

\subsubsection{Measurements in $\rm\bf B^+$ and $\rm\bf B^0$ decays}

Various other strategies, involving the decays of \bplus\ and \bo\ mesons,
have been proposed to extract or constrain $\gamma$. Some of these
have been studied for BaBar \cite{babarbook} and LHC-b \cite{lhcbtp}.
One method is to use
rare $\bplus\rightarrow\rm D^0K^{*+}$ and 
$\bo\rightarrow\rm D^0K^{*0}$ decays, 
however the combined branching ratios are so small that only a few events
per year are expected at BaBar, and a few tens of events at LHC-b. 
In $10^9$ \zb\ the linear collider should collect similar numbers to LHC-b,
so if these analyses are possible at all, the linear collider should be 
able to make a useful contribution. However, these channels have not yet
been studied in detail.

Another proposed method is to make use of flavour symmetries in 
$\bo\rightarrow\rm K^\pm\pi^\mp$ and $\bplus\rightarrow\rm K^0\pi^+$ decays.
The feasibility of reconstructing two body B decays has been discussed
in Section~\ref{s:alpha}. The reconstruction of $\rm K^0_s$ has not yet been
studied but should present no major problems. Again, if this analysis is
possible, the linear collider should be able to make a useful contribution.
However, the measurement of $\bplus\rightarrow\pi^+\pi^0$ is also required,
and if the penguin decay $\bplus\rightarrow\rm K^+\pi^0$ is also important,
the linear collider will be at a disadvantage in not having particle ID---the
mass resolution cannot be used effectively in final states containing 
reconstructed $\pi^0$.

\section{Summary and Conclusions}\label{s:conc}

The possibilities for doing electroweak and CP violation physics using a
sample of $10^9$ hadronic \zb\ decays 
collected at a linear collider \zb\ factory with
polarised beams have been investigated.
Huge improvements in the measurements of the \zb\
coupling parameters $\cAe$ and $\cAb$ will be possible. 
$\cAb$ can be measured to
a precision on $10^{-3}$ and $\cAe$ to a precision of $10^{-4}$, corresponding
to an error in $\swsqeffl$ of $0.000013$. However, to make this measurement 
useful the hadronic cross section below the $\Upsilon$ needs to be measured to
about 1\% precision, in order to calculate the running of the fine
structure constant up to the \zb\ mass.
If the Higgs has already been found, a stringent test of the Standard Model 
will be possible, and if not the Higgs mass can be predicted with 
5\% precision. Some interesting predictions in other models, such as the
MSSM, can also be made \cite{sven}.
However, the \zb\ partial width measurements cannot be improved very much
due to systematic error limitations.
For $\Rb$ a factor five may be possible, while for $\alpha_s(\MZ^2)$ and
$\Delta \rho$ only a factor two is possible.

For the CP violation studies,
whilst no `golden' measurements have been found that can be
performed better than at any other facility, many useful measurements can be
made that are complementary, and of a competitive precision, to those at 
$e^+e^-$ b factories, the Tevatron and LHC. These measurements
can be made with the standard linear collider detector optimised
for high energy running, and do not require dedicated particle
identification detectors.
Useful contributions can be made to the study of the CKM angles $\alpha$ and
$\beta$ in a variety of decay modes. However, the study of $\gamma$, which
mainly requires \bs\ meson decays, does not appear to be competitive with
the LHC, due to the much larger numbers of b decays produced at the latter
machine. 

In summary, with relatively little investment in time and money a large
improvement in the electroweak precision observables can be achieved, and
some competitive and complimentary CP violation measurements can be 
performed. The \zb\ factory option is thus an attractive  supplement 
to the already rich linear collider physics program.

\section*{Acknowledgements}
The authors would like to thank R.~Heuer, F.~Jegerlehner, T.~Mannel, 
T.~Ohl, P.~Rowson, M.~Winter and M.~Woods for useful discussions.

\end{document}